\documentclass[aps,twocolumn,prx,superscriptaddress]{revtex4-1}
\usepackage{graphicx}
\usepackage{dcolumn}
\usepackage{bm}
\usepackage[T1]{fontenc}
\usepackage[latin9]{inputenc}
\usepackage{textcomp}
\usepackage{amsmath}
\usepackage{mathrsfs}
\usepackage{graphicx}
\usepackage{amssymb}
\usepackage{units}
\usepackage{xcolor}
\usepackage{ulem}

\newcommand{\corr}[1]{{\color{black} #1 }}

\begin{document}

\title{Dispersion relation of the collective excitations in a resonantly driven polariton fluid}

\author{Petr Stepanov}
\affiliation{Univ. Grenoble Alpes, CNRS, Grenoble INP, Institut N\'{e}el, 38000 Grenoble, France}
\author{Ivan Amelio}
\affiliation{INO-CNR BEC Center and Dipartimento di Fisica, Universit\`a di Trento, 38123 Povo, Italy}
\author{Jean-Guy Rousset}
\affiliation{Institute of Experimental Physics, Faculty of Physics, University of Warsaw, Hoza 69, 02-681 Warszawa, Poland}
\affiliation{Univ. Grenoble Alpes, CNRS, Grenoble INP, Institut N\'{e}el, 38000 Grenoble, France}
\author{Jacqueline Bloch}
\affiliation{Centre de Nanosciences et de Nanotechnologies, CNRS, Universit\'{e} Paris-Sud, Universit\'{e} Paris-Saclay, C2N Marcoussis, F-91460 Marcoussis, France}
\author{Aristide Lema\^itre}
\affiliation{Centre de Nanosciences et de Nanotechnologies, CNRS, Universit\'{e} Paris-Sud, Universit\'{e} Paris-Saclay, C2N Marcoussis, F-91460 Marcoussis, France}
\author{Alberto Amo}
\affiliation{Univ. Lille, CNRS, Physique des Lasers Atomes et Mol\'ecules, F-59000 Lille, France}
\author{Anna Minguzzi}
\affiliation{Univ. Grenoble Alpes, CNRS, LPMMC, 38000 Grenoble, France}
\author{Iacopo Carusotto}
\affiliation{INO-CNR BEC Center and Dipartimento di Fisica, Universit\`a di Trento, 38123 Povo, Italy}
\author{Maxime Richard}
\affiliation{Univ. Grenoble Alpes, CNRS, Grenoble INP, Institut N\'{e}el, 38000 Grenoble, France}




\date{\today}

\maketitle

\textbf{Exciton-polaritons in semiconductor microcavities constitute the archetypal realization of a quantum fluid of light. Under coherent optical drive, remarkable effects such as superfluidity, dark solitons or the nucleation of hydrodynamic vortices have been observed. These phenomena can be all understood as a specific manifestation of collective excitations forming on top of the polariton condensate. In this work, we performed a Brillouin scattering experiment to measure their dispersion relation $\omega(\mathbf{k})$ directly. The result, such as a speed of sound which is apparently twice too low, cannot be explained upon considering the polariton condensate alone. In a combined theoretical and experimental analysis, we demonstrate that the presence of a reservoir of long-lived excitons interacting with polaritons has a dramatic influence on the nature and characteristic of the quantum fluid, and that it explains our measurement quantitatively. This work clarifies the role of such a reservoir in the different polariton hydrodynamics phenomena occurring under resonant optical drive. It also provides an unambiguous tool to determine the condensate-to-reservoir fraction in the quantum fluid, and sets an accurate framework to approach novel ideas for polariton-based quantum-optical applications.}

\begin{figure}[t!]
\includegraphics[width=0.9\columnwidth]{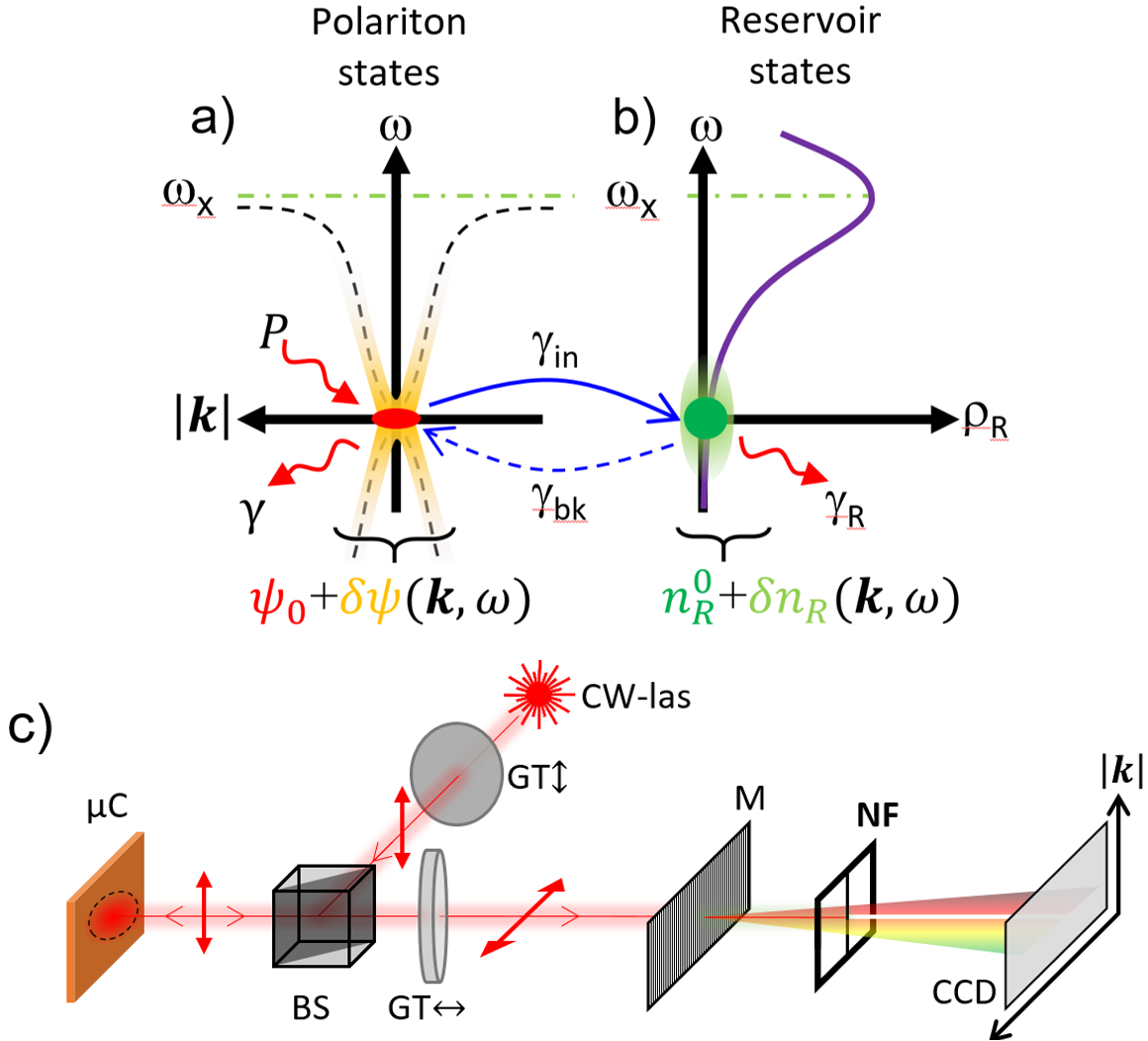}
\caption{\textbf{Illustration of the different fluid components and of the detection scheme} -  a) The polariton condensate (red spot) of wavefunction $\psi_0$, and of radiative loss rate $\gamma_c$ is resonantly excited by the laser of power P. An illustration of the condensate excitations $\delta\psi(\mathbf{k},\omega)$ are shown in yellow in the dispersion plane, with its typical dispersion relation (DR) $\omega(|\mathbf{k}|)$ shown in a black dashed line. The bare quantum well excitonic transition energy $\hbar\omega_X$ is shown as a green dashed line. An illustration of the typical quantum well excitonic density of state $\rho_R(\omega)$ is shown as a purple line in b). Owing to their effective mass differences, $\rho_R$'s peak value exceeds the polaritonic density-of-state by 5 orders of magnitude. The low energy tail of $\rho_R(\omega)$ originates from disorder in the quantum well, and can accommodate a reservoir (green spot) of long-lived excitons (loss rate $\gamma_R$, fluctuations $\delta n_R$ represented in light green). Interconversion of polaritons into reservoir excitons and back by optical absorption or scattering, occur at rates $\gamma_{in}$ and $\gamma_{bk}$ respectively. c) sketch of the experimental set-up used to measure the DR. The excitation laser light is linearly polarized by a Glan-Thompson polarizer (GT) and passed through a beam splitter (BS) to excite resonantly the polariton fluid. The cross-polarized reflected signal is selected by another GT and passed through a monochromator (M).The polariton emission at the laser frequency is further rejected by a metallic filter playing the role of notch filter (NF) and the remaining EPL is detected on a CCD camera. Some optical elements are omitted, that provide resolution on the EPL emission angle $\theta$, and thus on its in-plane wavevector $|\mathbf{k}|=(\omega/c)\sin(\theta)$ ($\hbar\omega$ is the photon energy, and $c$ the speed of light in vacuum).}

\label{fig1}
\end{figure}

\begin{figure*}[bt]
\includegraphics[width=0.9\textwidth]{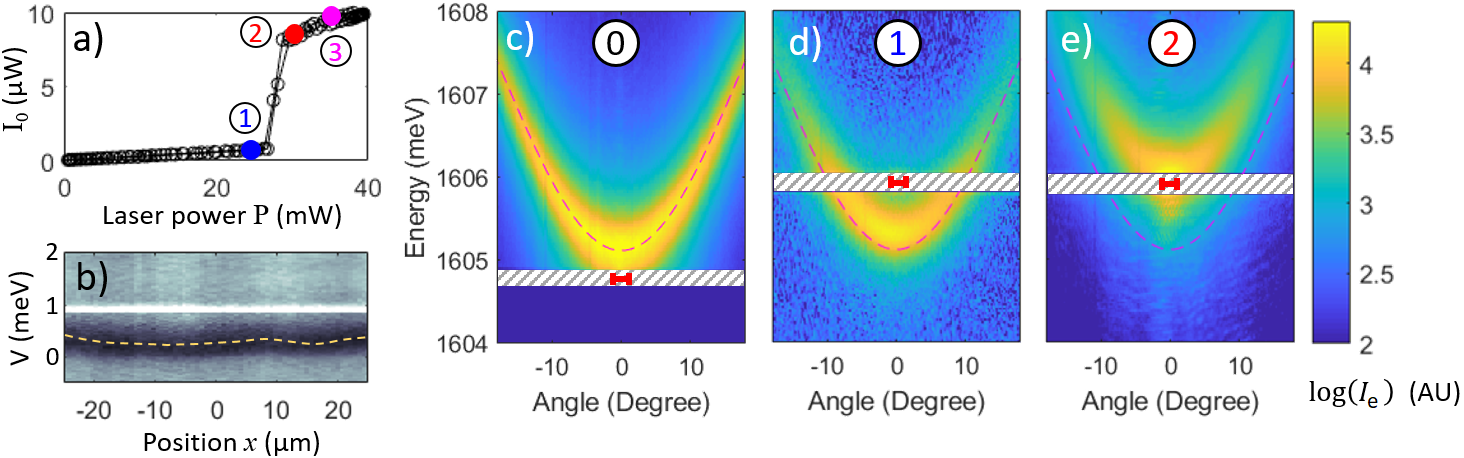}
\caption{\textbf{Characterization and Resonant photoluminescence of WPA} - a) Measured cross-polarized polaritonic emission intensity $I_0$ vs laser power $P$. b) Color-scaled microreflectivity measured across the working poin. The black area is the polariton absorption dip. The white streak is due to residual laser light. The orange dashed line show the extracted potential $V(x,y)$ across the working point. Angle and energy resolved cross-polarized EPL measurements $I_e(\theta_p,\hbar\omega)$ in the free-particle regime (c), in the lower branch of $I_0(P)$ (d), in the upper branch (e). The intensity is color coded on a logarithmic scale. The hatched rectangles show the spectral range rejected by the notch filter. The laser energy and angular spread is shown as a red segment. WPA is characterized by $\delta=+1.2\,5$meV}
\label{fig2}
\end{figure*}

Upon quieting down the thermal fluctuations in a liquid or a gaseous many-body system by deeply cooling it, and if it does not turn solid, a radical transformation occurs as the system behaviour starts to be dominated by quantum mechanics. In the case of integer spin particles, a so-called Bose-Einstein condensate appears below a critical temperature, in which a significant fraction of the fluid occupies a single quantum state \cite{huang_1963,pita_2016}. The system is then governed by the laws of quantum hydrodynamics in which the condensate phase $\phi(\mathbf{r},t)$ plays a central role, as well as the presence of two-body interaction. This framework explains key phenomena such as, irrotationality of the flow, quantization of vortex circulation, coupled amplitude-phase solitons, as well as the occurrence of a superfluid state. Among the most famous examples of such quantum fluids is superfluid helium \cite{kapitza_1938,allen_1938}, in which this regime has been pioneered, and ultracold atom gases \cite{dalfovo_1999} that are nowadays among the most accurate systems to investigate every subtleties of quantum fluids.

The root cause of these fascinating phenomena can be traced back to the nature and dispersion relation (DR) of the elementary excitations in the quantum fluid. As a general feature, these excitations consist of coupled phase and density fluctuations, and due to two-body interaction they are collective in nature. In the mean-field regime where the temperature is low, interactions are weak, and the condensate fraction is dominant, Bogoliubov derived in 1947 an analytical description of this regime, and could thus reveal the link between the DR of the excitations and the superfluid state \cite{bogoliubov_1947}. The corresponding experimental measurements of the elementary excitations came decades later in an ultracold atom gas \cite{raman_1999,chikkatur_2000,steinhauer_2002} and confirmed their key role in the observed phenomena.

\begin{figure*}[bt]
\includegraphics[width=0.75\textwidth]{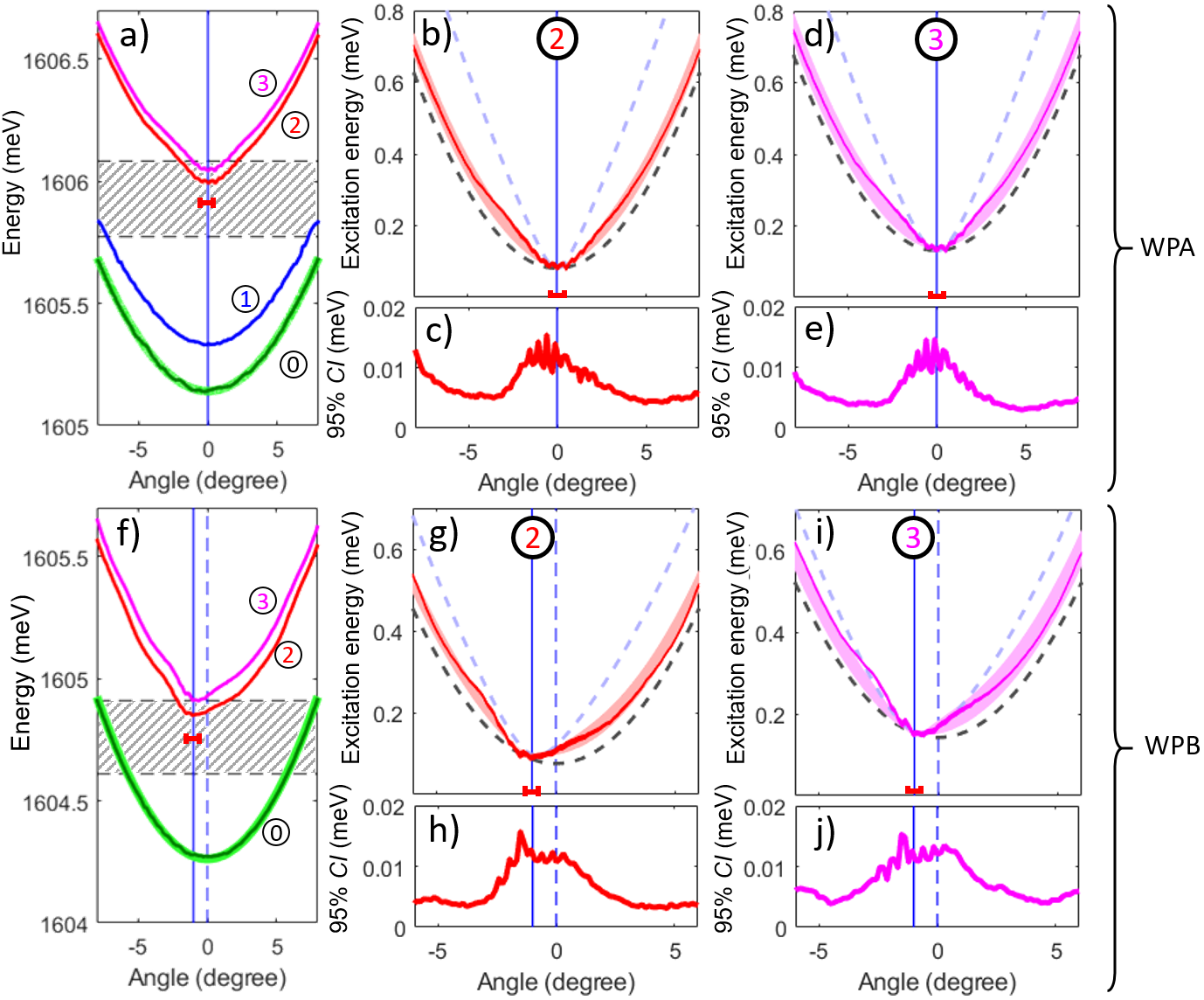}

\caption{\textbf{Excitations dispersion relations} - Measured DR $\hbar\omega(\theta)$ as obtained from the numerical analysis of the EPL (solid colored lines) for WPA (a) and WPB (f). Four states of the fluids are shown according to their position along $I_0(P)$ (plotted in Fig.\ref{fig2}.a for WPA): '0' is the free-particle DR (green line), '1' is a lower branch state of $I_0(P)$ (blue), '2' and '3' are two upper branch states (red and magenta). The calculated free polariton DR is the thick light green line underneath the measured one. b,d and g,i are zoomed in plots of DR '2' (narrow red line) and '3' (narrow magenta line) of  WPA and WPB respectively \corr{in units of condensate excitation energy $\hbar\omega_e=\hbar\omega-\hbar\omega_l$}. The $95\%$ confidence interval amplitude (CI) of the measured DRs is shown for each DRs in panels c,h and e,j for WPA and WPB respectively, with the same color codes. The theoretical DRs in the rigid blueshift limit (dashed black line) and in the GGPE limit (blue dashed line) are shown in b), d), g) and i). The full vectorial theory is also shown as a thick red and magenta lines visible underneath the measured DR; the line thickness represents the 95\% confidence interval of the fitting procedure with the data. In each panels, the laser energy and angular spread is shown as a red segment, and the vertical blue line shows the angle of excitation. The hatched rectangle is the spectral range rejected by the notch filter in a) and f).}
\label{fig3}
\end{figure*}

Recently, quantum fluids of light have emerged as a new class of quantum fluids, characterized by their nonequilibrium character~\cite{carusotto_2013,klaers_2010}. A paradigmatic member of this class is the fluid of exciton-polaritons (polaritons), that can be pumped within the spacer of an optical microcavity in the strong coupling regime~\cite{weisbuch_1992}. Polaritons can be understood as photons dressed by semiconductor electrons-hole pairs (excitons), that display significant binary interactions mediated by Coulomb interaction. Their nonequilibrium character comes from the fact that they need to be continuously replenished by an external pump to compensate for ultra-fast radiative losses. They thus require a different theoretical framework than their equilibrium counterpart, which is based on the generalized Gross-Pitaevskii equation (GGPE)~\cite{carusotto_2004}.

In this nonequilibrium setting, the notion of superfluidity raises intriguing questions about its definition and characteristic observables ~\cite{wouters_2010,keeling_2011,janot_2013,regemortel_2014,gladilin_2016,juggins_2018}; experimentally, a superfluid-like frictionless flow has been demonstrated in 2009 in a steady-state polariton fluid~\cite{amo_2009}. From there on, a number of quantum hydrodynamics phenomena have been studied in this system, including the nucleation of topological excitations such as solitons~\cite{amo_2011,sich_2012} and quantized vortices~\cite{lagoudakis_2008,nardin_2011,sanvitto_2010}, phenomena associated with their spin degree of freedom~\cite{hivet_2012,lagoudakis_2009}, vortex dynamics~\cite{lagoudakis_2011,dominici_2018}, and turbulence effects~\cite{grosso_2011}.

As mentioned above, the DR of the collective excitations play a crucial role in these phenomena. To the best of our knowledge, while it has been investigated in some particular cases, its measurement under resonant optical drive is still missing. In \cite{kohnle_2011} for instance, a transient polariton population was created with a resonant pulsed laser and the time-integrated DR of the excitations was measured in a four-wave mixing arrangement, hinting at sound-like collective excitations. A measurement under resonant continuous wave (CW) drive has been carried out in~\cite{zajac_2015}, where two bands of positive and negative frequencies excitations have been measured, and found in agreement with the normal and ghost branches of Bogoliubov's theory. However, the optical pump was kept too low to observe collective features beyond the perturbative regime of the single particle picture. The DR of collective excitations has also been measured in the regime of pulsed non-resonant optical drive in several works \cite{yamamoto_2009,assmann_2011,pieczarka_2015}, but in this incoherent excitation configuration, the nature and DR of the excitations are different from those realized under resonant drive condition like ours~\cite{wouters_2007}. Moreover, the pulsed excitation introduces time-dependent densities, which results in inhomogeneous broadening of the excitation spectrum, and makes comparison with theory harder. Finally, a measurement of the Bogoliubov DR has been reported recently in an equilibrium-like fluid of light in the conceptually different case of a propagating geometry in an atomic vapor cell \cite{fontaine_2018}.

In this work, we resonantly drive a \corr{nonequilibrium condensate of exciton-polaritons with a CW laser, in which the long-range coherence is directly imprinted by the laser, and not the result of a condensation mechanism. We} focus our attention \corr{on the high density regime in which the interaction energy is comparable or exceed the linewidth, such that the excitations are of collective nature}. We perform a direct measurement of their DR using an angle-resolved spectroscopy technique inspired from Brillouin scattering experiments. We find that the results differ strongly from the pure polariton condensate situation described by the GGPE, like for example a speed of sound which is apparently twice too low. Inspired by previous work suggesting that a reservoir of long-lived excitons coexists with polaritons even in this resonant excitation regime \cite{paraiso_2010,wouters_2013,vish_2012,sekretenko_2013,krizha_2017,baboux_2018}, we took this possibility into account in a novel theoretical framework in which polaritons can be converted into reservoir excitons (cf. illustration in Fig.\ref{fig1}.a,b), and in which the reservoir provides an additional two-body interaction channel. The resulting excitations are of hybrid reservoir-density and Bogoliubov-excitations nature, and agree quantitatively with our measurements. \corr{While some qualitative analogy may be found with second sound in liquid Helium~\cite{pita_2016}, there are major differences due to the non-equilibrium nature of the quantum fluid; moreover, this hybrid nature results in} quantitative corrections with respect to the GGPE description, of importance for both past and future works on polaritons quantum hydrodynamics.

\section{EXPERIMENTAL RESULTS}

The experiment is carried out with a liquid Helium cooled planar GaAs/AlAs microcavity in the strong coupling regime. The coherent polariton fluid (referred to thereafter as 'the condensate') is excited resonantly with a tunable single longitudinal mode CW laser (cf. Methods). A sizable population of excitations is spontaneously created on top of the resonantly driven polariton condensate by the interaction of polaritons with the thermal bath of acoustic phonons naturally present within the solid-state microcavity (see SI section III). The polaritons involved in these excitations can then relax radiatively, so that their energy and momentum with respect to the condensate constitute a direct measurement of the DR of the excitations. This emission, that we will refer to as excitations photoluminescence (EPL), is collected by the detection scheme illustrated in Fig.\ref{fig1}.c.

The collective excitations that differ from free-particle excitations live within a small frequency window surrounding the condensate, of width comparable with the interaction frequency $g n \gtrsim 0.5\,$meV/$\hbar$, where $g$ and $n$ are the polariton-polariton interaction constant and the condensate density respectively. We thus isolate the EPL $I_e$ from the much brighter condensate intensity $I_0$, and from the Rayleigh scattered laser light using a two-stage filtering scheme: in the first stage, we profit from the fact that the condensate has a nonzero cross-polarized component caused by a weak residual birefringence and a weak TE-TM splitting~\cite{panzarini_1999}, to detect the EPL in the cross-polarized direction with respect to the laser. For the second stage, we designed an narrow band notch filter (labelled 'NF' in Fig.\ref{fig1}.c) made up of a featureless metallic stripe placed in the output focal plane of the monochromator. The resulting rejection is such that the EPL signal can be well identified even as close as $0.1\,$meV from the condensate. Fig.\ref{fig2}.c-e show measured angle-resolved EPL patterns $I_e(\theta,\omega)$ obtained with this method.

In order to measure sharply-defined dispersion relations, we chose two regions of the microcavity of $\sim 50\mu$m diameter characterized by a weak disorder amplitude of the potential experienced by polaritons $V(x,y)$. They are labelled further on as 'working points' (WPs) A and B (For the sake of generality, a third working point is presented and fully analyzed in the SI section IV). The microreflectivity measurement shown in Fig.\ref{fig2}.b provides a cross-section of $V$ across WPA showing that its spatial fluctuations are smooth and small as compared to the linewidth. The reference free-particle DR at WPA is extracted from the EPL measurement shown in Fig.\ref{fig2}.c. It is obtained under weak excitation at normal incidence, with the laser energy $\hbar\omega_{l}$ red-detuned from the $\mathbf{k}=0$ lower polariton resonance by $\Delta=\hbar\omega_{l}-\hbar\omega(0)=-0.5\,$meV. The corresponding extracted free-particle DR is labeled '0' in Fig.\ref{fig3}.a.

We then shifted the laser to $\Delta=0.79\,$meV on the blue side of the polariton resonance, in order to access the regime for which $n(P)$ the condensate density dependence on the driving laser power $P$ exhibits a bistable behaviour~\cite{baas_2004,boulier_2014}. In the context of the GGPE theory, the regime of collective excitations corresponds to the upper branch of the bistable $n(P)$: at the lower laser power edge of this branch (just before switch down) sits the gapless sonic regime, in which the excitations are expected to be phonon-like with a well-defined speed of sound~\cite{carusotto_2013}. Higher up along this branch, a gap opens up for increasing $P$ and the DR adopts a more curved shape. In order to characterize this bistability curve, the unfiltered condensate emission $I_0$ is collected in the cross-polarized direction vs the laser power $P$. Note that in this measurement the excitations have a negligible contribution. The measured $I_0(P)$ curve is shown in Fig.\ref{fig2}.a: the lower and upper branch are separated by a sharp jump: indeed since we chop the laser with a $5\%$ duty-cycle to prevent unwanted heating effects, the bistable region appears closed. Note that in spite of this technical constraint, we can still get very close to the sonic regime in the measurements, although we can never strictly reach it.

Based on this preliminary calibration measurement, we proceeded to the extraction of the DR of the collective excitations for several laser powers $P$ along the upper branch $I_0(P)$. Due to the stringent requirements of the driving laser beam shape both in Fourier and real space, we chose to work with a large laser spot of Gaussian intensity profile of $50\,\mu$m diameter (cf. Methods). As a result, for states in the upper branch of $I_0(P)$, the polariton density is organized into two large spatial structures: The high polariton density is contained in a large diameter disk-shaped area at the center of the laser spot, separated from a low density outer region by a sharp switching front (see examples in the SI, Fig.S5). \corr{The nonlinearity thus acts as an effective 'top-hat' spatial filter for the Gaussian pump mode, that homogenize the high-density region we are interested in (see details in SI section V.3). We also checked in a lineshape analysis that the influence of the in-plane disorder visible in Fig.\ref{fig2}.b is weak and negligible as compared to the features of interest in the dispersion relation (cf. SI section V.4).}

In order to collect EPL only from the high density area, we rejected the outer region using an iris of diameter $D_i$ matching that of the switching front \corr{of typically $\sim 35\,\mu$m diameter. While this spatial selection introduces a spurious angular spread $\leq 1.5^\circ$ to the EPL, it does not prevent resolving the collective features in the dispersion relation, that are visible within a $\sim 5^\circ$ window as can be seen in Fig.\ref{fig3}. As discussed in section VIII in the SI, we took this finite angular resolution into account in the simulations. Note that under non-resonant incoherent excitation, another source of momentum broadening would appear due to density-induced radial flow of polaritons \cite{wouters_2008,richard_2005}. This cannot occur in our resonant excitation configuration since the condensate phase is locked by the coherent pump.}

Two raw results of angle- and energy-resolved EPL are shown in Fig.\ref{fig2}.d and Fig.\ref{fig2}.e that correspond to two states on the curve $I_0(P)$ labeled '1' and '2' respectively in Fig.\ref{fig2}.a: '1' is a state on the lower branch of $I_0(P)$, while '2' is on the upper branch and as close as possible to the switch-down point.

The key feature we want to focus on in this work is the shape of the collective excitations DR $\hbar\omega_e(\theta)$ (where $\hbar\omega_e=\hbar\omega-\hbar\omega_l$, and $\hbar\omega$ is the detected photon energy), and how it compares with common assumptions. We thus performed a numerical analysis of the raw angle and energy-resolved EPL measurements in order to determine as accurately as possible the measured DR, with a special care taken on determining the statistical confidence interval of the result (see Methods section). Fig.\ref{fig3}.a show the resulting DRs for WPA, for the free polariton dispersion (dark green line, labelled '0'), and three different blueshifted states (blue, red and magenta lines, labelled '1' to '3'), with a detailed view of '2' and '3' shown in Fig.\ref{fig3}.b and Fig.\ref{fig3}.d respectively, and the 95\% confidence interval for the determination of $\hbar\omega_e$ shown in Fig.\ref{fig3}.c and Fig.\ref{fig3}.e respectively. The same analysis is shown for WPB in Fig.\ref{fig3}.f-j (see SI section II for details on the method). The free polariton DRs in WPA and WPB are very well-fitted with the near-parabolic theoretical free polariton DR (light green line) as obtained from the coupled oscillators model, and can thus be trusted for comparison purpose.

We first focus on WPA for which the laser drive is at normal incidence: Curve '1' shows the DR of a polariton fluid for which the blueshift with respect to DR '0' amounts to  $\hbar\omega_{BS}=0.18\,$meV, and which is still on the lower branch of $I_0(P)$. Its shape is identical to '0', indicating that in spite of the blueshift, the condensate excitations are only weakly perturbed from the free-particle picture. Curves '2' and '3' are obtained in the upper branch of $I_0(P)$ and feature a clearly modified shape with respect to '0', which is an unambiguous signature that the nature of the condensate excitations have changed from free-particle to collective as a result of strong interaction energy ('2' and '3' are blueshifted by $\hbar\omega_{BS}=0.85\,$meV and $\hbar\omega_{BS}=0.90\,$meV respectively, with respect to '0'). In Fig.\ref{fig3}.b and \ref{fig3}.g, the low-energy low-angle part of both DR are compared with the theoretical shape expected in two limiting situations: (i) the condensate density is small as compared to the reservoir density, such that the DR consists of a rigidly blue-shifted free-particle dispersion; (ii) The system consists of a pure condensate, without any reservoir fraction so that the DR has the form given by the GGPE.

In mathematical form, the DR in case (i) reads
\begin{equation}
\omega_{RB}(\mathbf{k})=\omega_{0}(\mathbf{k})+g_R n_R-i\,\frac{\gamma_{c}}{2},
\label{eq:freepart}
\end{equation}
where $\gamma_{c}$ is the polariton radiative loss rate, $\hbar\omega_{0}(\mathbf{k})$ is known from the DR measurement at point '0', and $n_R$ and $g_R$ are the reservoir particles density and their interaction constant with the condensate respectively . $\hbar\omega_{RB}(\theta)$ is plotted as a black dashed line in Fig.\ref{fig3}.b and g. In this model, the interaction term is fixed directly by the blueshift as $\hbar g_Rn_R=\hbar\omega_{BS}$. The comparison between this model and the measured dispersion '2' and '3' show a clear mismatch, in which the measured dispersion is steeper. The theoretical shape of the dispersion in case (ii) can be obtained by linearizing the GGPE \cite{carusotto_2004} around the pure condensate steady-state density, which, for a condensate with zero momentum results in:
\begin{multline}
\omega_{\rm Bog}(\mathbf{k})= \omega_{p}
\pm \sqrt{(\omega_0(\mathbf{k}) +2 g n -\omega_{p} )^2  - (g n)^2}-i\,\frac{\gamma_{c}}{2}.
\label{eq:bogo}
\end{multline}
Like in the previous case, the interaction energy $\hbar gn$, where $g$ is the interaction constant and $n$ the polariton density, is inferred unambiguously from the experimentally measured blueshift $\hbar gn=\hbar\omega_{BS}$, which fixes also the position on the theoretical curve $I_0(P)$. These calculated DRs are shown in Fig.\ref{fig3}.b and g as dashed light blue lines. This time, both for '2' and '3', the measured dispersions are now clearly not steep enough to match the theory.

We then performed the same analysis for the other working point WPB, in which we used a nonzero laser incidence angle $\theta_p=-1^\circ$. In WPB, owing the microcavity tuning, the interactions are smaller by a factor $\sim 2$ and a smaller laser detuning of $\Delta=0.47\;$meV is chosen accordingly. The local disorder is obviously different, but of similar average amplitude and characteristic length, as in WPA. The measured DRs for WPB are shown in Fig.\ref{fig3}.f-j with the same labelling conventions as for WPA. For dispersions '2' and '3', situated on the upper branch of $I_0(P)$, with '2' as close as possible to the switch down point, an asymmetric shape of the DR is obtained, as expected for the collective excitations when the condensate is subject to a flow of nonzero velocity.

Like in WPA, the comparison with the two theoretical limiting cases are shown in Fig.\ref{fig3}.d,i and demonstrates that the measured DRs do not agree with either of them. Our analysis also gives us access to the spectral linewidth $\hbar\gamma$ of the excitations as a function of $k$. The latter is found to be essentially fixed by the radiative linewidth, plus an additional contribution coming from the propagation time within the finite size polariton fluid. A more detailed discussion can be found in the SI section V.2.

\section{THEORY AND INTERPRETATION}

These experimental observations pinpoint the necessity of building a theoretical description of the elementary excitations that includes a contribution from the reservoir. We first check the existence of the reservoir alongside polaritons by exploiting their very different respective lifetimes: we performed a time-resolved polariton photoluminescence decay measurement under pulsed resonant excitation and found that the polariton population exhibits a fast decay component fixed by the polariton lifetime, and a slow decay time component of $400\,$ps that indicates clearly the formation of a reservoir of long-lived excitons fed by the polaritons. Our best guess considering the experimental indications at hand, is that these low energy excitonic states are due to disorder in the quantum well \cite{savona_2007}. A detailed discussion can be found in the SI section I.

We thus developed a theoretical model with the GGPE for the condensate wavefunction $\psi$ as a starting point, coupled to a rate equation that describes the reservoir density dynamics $n_R(t)$~\cite{wouters_2007}. In addition, due to the cross-polarized measurement technique, this theory needs to account for the polariton polarization degree of freedom. This vectorial theory is presented in details in the SI section VIII. For the sake of simplicity, we present thereafter a scalar version in which the coupled equations read
\begin{eqnarray}
  i\hbar \partial_t \psi&=&\left[\hbar \omega_0 -\frac{\hbar^2}{2m} \nabla^2 + \hbar g|\psi|^2 + \hbar g_R n_R\right.  \nonumber
  \\ &-&\left.i \frac{\hbar (\gamma_c+\gamma_{in})}{2}\right]\psi +F(t)  \label{eq:cond} \\
  \partial_t n_R &=&\corr{-\gamma_R\, n_R +\gamma_{in}|\psi|^2},
  \label{eq:reserv}
\end{eqnarray}
where the condensed polaritons of effective mass $m$ are resonantly and coherently driven by a homogeneous pump $F(t)$, their finite lifetime is limited by the radiative loss rate $\gamma_c$, and their \corr{(typically much slower)} capture rate by the reservoir is $\gamma_{in}$. The polariton-polariton interaction energy is proportional to the density $n=|\psi|^2$ with a coupling constant $g$, while the interactions between polaritons and the reservoir population contribute in an additional interaction energy $\hbar g_R n_R$. The reservoir population lifetime is fixed by the time-resolved experiment to $400\,$ps, such that $\hbar\gamma_R=1.6\,\mu$eV $\ll \hbar\gamma_c$. \corr{As illustrated in Fig.\ref{fig1}.a,b and explained in the SI section I, the reservoir decay also involves processes in which the reservoir excitons are converted back into non-condensed polaritons. In eq.(\ref{eq:cond}), this latter process would result into a stochastic source term for the excitations of the fluid~\cite{wouters_2009}. Experimentally, we find that the relative intensity of the excitations is very small as compared to the coherent field, so that a linearized treatment around the steady-state is an accurate description.}

\corr{This model describes a condensate on top of which excitations of frequencies $\omega_e(k)/2\pi$ are present, that can be classified into two distinct regimes. On one hand, the excitations of zero-frequency $\omega_e=0$ correspond to the formation of a static spatial pattern as generated e.g. by a condensate flowing against an obstacle. It is governed by the total population of the condensate plus reservoir, i.e. the total blueshift $\hbar\omega_{\rm BS}=\hbar g|\psi|^2+\hbar g_Rn_R$. In this limit, our model can be mapped onto a GGPE, in which an effective interaction term defined as $g_{\rm eff,s}=g+g_R\gamma_{in}/\gamma_R$ such that $g_{\rm eff,s}n=\hbar\omega_{\rm BS}$ is used, and contains the reservoir contribution. Then, the known properties of the GGPE in the static limit \cite{carusotto_2004} do apply. In particular, the suppression of polariton scattering due to interaction with an obstacle is expected to happen below a critical velocity $v_c=\sqrt{\hbar\omega_{\rm BS}/m}$, which is in agreement with the pioneering experimental reports of superfluidity \cite{amo_2009, amo_2011}. On the other hand, in the regime where excitations have a frequency $\omega_e/2\pi\gg\gamma_R$, as is the case in this work, the reservoir response is too slow to follow the condensate fluctuations. Our model can still be mapped onto a GGPE but with a different effective interaction $g_{\rm eff,d}=g$, and $g_{\rm eff,d}n<\hbar\omega_{\rm BS}$. In this non-zero frequency limit, the speed of sound $c_s$ of the gapless sonic state is well-defined and is governed solely by the polariton condensate fraction: $c_s \simeq \sqrt{\hbar gn/m}$. Note that for this particular sonic state, $c_s<v_c$, which means that the standard prediction of Landau's critical velocity must be modified. This is due to the fact that excitons in the reservoir are fixed in the material frame, which in turn affects the Galilean relativity features of the coupled system. Based on these arguments, the contributions to the blue shift due to polariton-polariton and polariton-reservoir interactions can be separated by comparing the polariton interaction extracted from the measured speed of sound $\hbar gn=m c_s^2$ and the total blue shift $\hbar\omega_{BS}$.}

In order to simulate the experimental measurements, we used the vectorial version of the model, and determined the energy-momentum power spectrum of the condensate wavefunction by convolving the Bogoliubov matrix response function with a stochastic Gaussian noise simulating the coupling with the phonon bath. We then extracted the DR by fitting these data using the same method as for the experimental data. We obtained a quantitative agreement with the experimental DR both in WPA (thick red/magenta line in Fig.\ref{fig3}.b/d) and WPB (thick red/magenta line in Fig.\ref{fig3}.g/i). \corr{The only fitting parameter is the steady-state ratio of condensate to reservoir interaction energy}. Note that in this vectorial theory, the DRs exhibits a total of five branches: two on the positive energy side, two on the negative side, and a flat one that we do not have access to in this experiment (\corr{cf. section VIII in the SI}). As a result, the calculated DR that we use to fit our measurements consists of a non-trivially weighted contribution of two branches originating from the co- and crossed-polarized condensate, which also contribute to the linear part of the dispersion.

The best fit is obtained for \corr{$\rho=n/(n_R+n)=46\% [38\%;53\%]$ for WPA, and $73\% [60\%;85\%]$ for WPB, where we used $g_R=g/|X|^2$, where $|X|^2=0.58$ and $|X|^2=0.38$ for WPA and WPB respectively, $n=|\psi_x|^2+|\psi_y|^2$ and the indices ${x,y}$ refer to the co- and cross- polarized components respectively, and the bracket is the 95\%-confidence interval. The line thickness of the calculated DRs in Fig.\ref{fig3}.b,d (WPA) and in Fig.\ref{fig3}.g,i (WPB) show the area overlapped by the theoretical DRs within this interval. In the limit of vanishing birefringence, $|\psi_y|^2\ll|\psi_x|^2$, the characteristic speed of sounds amount to $c_x/v_c=0.42\pm0.05$ for WPA and $c_x/v_c=0.56\pm0.12$ for WPB (see details in the SI section VI). In addition, we can use this knowledge of the condensate contribution to the total blueshift to obtain a new estimate of the polaritons-polariton interaction constant, excluding interaction with the reservoir. Upon normalizing by the number of quantum well, we thus find $g_T=8\pm 2\,\mu$eV.$\mu$m$^2$ in WPA, which is mostly in-line the values previously reported in the literature \cite{delteil_2019, munoz_2019, krizha_2017}. Note that this value is to be considered as an order of magnitude as the experiment was not optimized for this quantitative derivation of $g_T$ (cf section VII in the SI for more details).}

\begin{figure}[bt]
\includegraphics[width=\columnwidth]{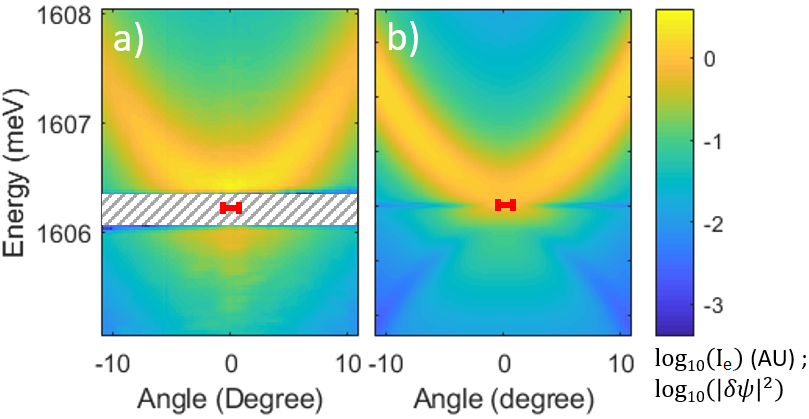}
\caption{\textbf{Superfluid excitations density analysis -} a) Measured cross-polarized EPL measurements in the upper branch of $I_0(P)$ close to the switch-down point, with an enhanced signal-to-noise ratio of 300. b) Calculated angle and energy-resolved cross-polarized EPL from the full vectorial theory assuming a condensate fraction of $\rho=50\%$ and experimental parameters from WPA. The intensity is color coded on a 4 decades logarithmic scale, and the laser energy and angular spread is shown as a red segment. The hatched rectangle show the spectral range rejected by the notch filter.}
\label{fig2p5}
\end{figure}

Remarkably, our model also recovers the fact that in our measurements, the ghost branch emission is strongly suppressed (cf e.g. Fig.\ref{fig2}.d,e). We double-checked this feature in the measurement shown in Fig.\ref{fig2p5}.a in which the signal-to-noise was enhanced to 300, and in which emission from the ghost branch is not found either. Our model explains the two different origins contributing to this suppression: (i) the ghost branch contribution vanishes when the incoherent (reservoir) fraction of the fluid increases, an effect which is common to the scalar situation. (ii) In cross-polarized detection geometry, the different polarization contributions to the excitations interfere in a non-trivial way as is visible in eq.(S16) in the SI section VIII. This interference can either enhance or, as in our case, suppress the ghost branch emission. \corr{Note that this mechanism is not directly related to the one discussed in~\cite{Hanai_2018} which refers to a different pumping scheme. Note also that} in \cite{zajac_2015}, special care was paid to reduce the excitonic disorder (i.e. the reservoir contribution) \cite{borri_2000,savona_2007}, such that the emission from the ghost branch could be detected, albeit in a regime dominated by the free particle properties.


\section{CONCLUSION}

In summary, we obtained a direct measurement of the dispersion relation of the collective excitations on top of a coherently driven polariton condensate. This measurement can be directly used as the key experimental ingredient required to test the system against most superfluidity criteria. Our analysis also provides a deep insight into the unique two-components nature of this quantum fluid and its elementary excitations: partly polariton condensate and partly incoherent reservoir. In particular, we have checked that the speed of sound, which is a key characteristic of collective excitations in quantum fluids, is strongly reduced by the presence of the reservoir. More generally, we have established theoretically and experimentally that the reservoir involvement results in two different regimes of collective excitations. On one hand, the spatial density patterns resulting from static collective excitations like in historical frictionless flow experiments~\cite{amo_2009,amo_2011}, involve the total blueshift due to polaritons and reservoir. On the other hand, finite-frequency collective excitations as investigated here, are governed only by interactions between condensate polaritons.

We are currently investigating the extent and magnitude of these consequences from both the theoretical and the experimental point of view. For instance, the hydrodynamic nucleation of vortices studied in~\cite{nardin_2011,pigeon_2011,pigeon_2017,amelio_2019} is a time-dependent phenomenon, for which one can naturally expect strong modifications due to the coupling of the coherent polariton fluid to the much slower degrees of freedom of the reservoir that strongly affects its Galilean relativity properties. From a quantum optical perspective, the additional blueshift of the polariton state due to the incoherent reservoir may be playing a role in the experimental recent two-body correlation experiments~\cite{delteil_2019,munoz_2019}. Along the same lines, the presence of this reservoir may have a dramatic impact on the accuracy of recent measurements of the polariton-polariton interaction constant~\cite{snoke_2017,estrecho_2018}. In this context, our spectroscopic measurement of the collective excitations provides an new and accurate way to extract the polariton-polariton interactions contribution to the total interaction energy, and isolate it from the reservoir contribution.

\section*{ACKNOWLEDGEMENTS}
The authors wish to thank M. Wouters for crucial discussions in the early stage of this project. P.S, J.B., A.A.G, and M.R. acknowledge funding from the french ANR contract "QFL" (ANR-16-CE30-0021). I.C. J.B. A.L. A.A. acknowledge funding from the H2020-FETFLAG-2018-2020 project "PhoQuS", (nb 820392). A.M. acknowledges funding from the french ANR contract "SuperRing" (ANR-15-CE30-0012). J.G.R. acknowledges the "ETIUDA" program from the polish NCN.

\corr{\section*{Data availability}
The data that support the findings of this study are available from the corresponding author upon reasonable request.}

\section*{Methods}

\subsection{Method: detail of the microcavity}
The GaAs/AlAs microcavity used in this experiment is identical to that used in \cite{bajoni_2007}. It features a quality factor $Q=3000$. \corr{This relatively low quality was chosen on purpose as it satisfies two conflicting requirements: (i) the need to have a collective excitation energy window that largely exceeds the instrumental resolution of $70\,\mu$eV, and (ii) the need to keep the laser intensity low, which requires the linewidth to be not too much smaller than $\Delta$.} The heavy-hole and light-hole excitonic transitions energy are at $E_{hh}=1612.05\,$meV and $E_{lh}=1644\,$meV at $T=30\,$K, and the corresponding Rabi splittings resulting from the strong coupling regime with the cavity mode of energy $E_{c0}$ are $\hbar\Omega_{hh}=15\,$meV and $\hbar\Omega_{lh}=12.5\,$meV respectively. \corr{the microcavity is intentionally wedged in order to tune $E_{c0}$ with respect to $E_{hh}$}. The background index of the microcavity is $n_{bg}=3.65$. The microcavity is placed in the vacuum chamber of a temperature-tunable Helium flux cryostat. The temperature is set at $T=30K$, found as an optimum of thermal phonons population and polariton linewidth. WPA (WPB) is characterized by a detuning $\delta=E_{c0}-E_{hh}=+1.25\,$meV ($\delta=-1.82\;$meV). The lower polariton mode exhibits a full width at half maximum of $\hbar\gamma_c=0.4$meV, and a reservoir decay energy $\hbar\gamma_R=1.6\,\mu$eV in both WPs.

\subsection{Method: Parameters of the resonant laser excitation}
The polariton fluid is driven resonantly with a single longitudinal mode CW Ti-Sapphire laser of $5\,$MHz linewidth, and linearly polarized with a high purity. The laser beam is shaped with pinholes into a spatially Fourier-limited Gaussian mode of $50\,\mu$m diameter as measured on the surface of the microcavity, and a corresponding $\delta\theta=\pm 2^\circ$ angular spread (i.e. $\delta k_\parallel=\pm 0.28\,\mu$m$^{-1}$) in momentum space. WPA is excited at normal incidence, while WPB is excited with a $-1^\circ$ incidence angle. The detuning $\Delta$ between the polariton mode and the CW laser are $\Delta=0.79\,$meV (WPA), and $\Delta=0.47\,$meV (WPB).

\subsection{Method: Full vectorial theory}
Owing to its polarization degree of freedom, the condensate field is a vectorial field, characterized by its two components $\psi_\sigma({\mathbf r})$ in the $\sigma={x,y}$. Each of these component is described by a GGPE equation coupled with the other, and additionally coupled to the dark-exciton reservoir via its density $n_R$:
\begin{eqnarray}
i\partial_t \psi_x &=& \left[ \omega_{LP} (\hat{\mathbf{k}}) - \frac{\alpha}{2} \cos(2\Theta) + \frac{g_T+g_S}{2} |\psi_x|^2 + g_T |\psi_y|^2 \right. \nonumber \\
&+& \left. g_R n_R  -i \frac{\gamma_c + \gamma_{in}}{2} \right] \psi_x - \frac{\alpha}{2} \sin(2\Theta) \, \psi_y \nonumber \\
&-& \frac{g_T - g_S}{2}\psi_x^* \psi_y^2 +  F
\end{eqnarray}
\begin{eqnarray}
i\partial_t \psi_y &=& \left[ \omega_{LP} (\hat{\mathbf{k}}) + \frac{\alpha}{2} \cos(2\Theta) + \frac{g_T+g_S}{2} |\psi_y|^2 +  g_T |\psi_x|^2  \right. \nonumber \\
&+& \left.g_R n_R  -i \frac{\gamma_c  + \gamma_{in}}{2} \right] \psi_y - \frac{\alpha}{2} \sin(2\Theta) \, \psi_x \nonumber \\
&-& \frac{g_T - g_S}{2}\psi_y^* \psi_x^2
\end{eqnarray}
\begin{equation}
\partial_t n_R = - \gamma_R n_R + \gamma_{in} (|\psi_x|^2 + |\psi_y|^2),
\end{equation}
\\
where  $\omega^{LP} (\hat{\mathbf{k}}) = \omega_0^{LP} -\frac{\hbar}{2 m} \mathbf{\nabla}^2 $ and $\hat{\mathbf{k}} = -i \nabla$, with $m$ the effective polariton mass,
$\alpha \sim 0.1 \pm 0.05$ meV is the birefringence splitting measured at normal incidence, and $\Theta \simeq 19^{\circ}$ is the angle between the birefringence axes and the cross-polarized measurement basis $x,y$.
$g_T$ and $g_S$ are  the triplet and singlet coupling constants respectively, and we take  $g_S= -0.1 \ g_T$ \corr{with $g_T>0$. The measurement of the steady-state dispersion relation provides us with a measurement of the condensate interaction energy $\bar{g}(|\psi_x|^2 + |\psi_y|^2)$, and on the reservoir interaction energy $g_Rn_R=g_R(|\psi_x|^2 + |\psi_y|^2)\gamma_{in}/\gamma_R$. An independent knowledge of $\bar{g}$, $g_R$ or $(|\psi_x|^2 + |\psi_y|^2)$ requires additional assumptions or measurements, like the exact impinging drive power at the hysteresis switch up point.} The pump field is $F({\mathbf r}, t) = F_0 e^{i {\mathbf k}_p\cdot {\mathbf r}-i\omega_p t}$. The other parameters are the reservoir filling rate $\gamma_{in}$ from the condensate, the reservoir decay rate $\gamma_R$ and the condensate radiative loss rate $\gamma_c$. The properties derived from this theory are discussed in details in the SI section VIII.

\subsection{Method: determination of the experimental dispersion relation and its confidence interval}
The experimental dispersion relations are extracted from the measurements of $I_e(k,\omega)$: For each column $i$ of $I_e(k_i,\omega_j)$ that contains the spectrum at wavevector $k_i$, the EPL emission peak is fitted with a Lorentzian peak. From this fit and its goodness, we get the central frequency $\omega_{0,i}/2\pi$ of the peak, and its $95\%$-confidence interval $\delta\omega_i $ respectively. The thus obtained ensemble $\omega_{0,i}(k_i)$ is the extracted dispersion relation, and ${|\delta\omega_i(k_i)|}_i$ is the $95\%$-confidence interval computed from the optimization algorithm used in the fitting procedure. This procedure is described step by step for WPA  point '3', in the SI section II.

\end{document}


\title{SUPPLEMENTARY INFORMATION: Dispersion relation of the collective excitations in a resonantly driven polariton fluid}

\author{Petr Stepanov}
\affiliation{Univ. Grenoble Alpes, CNRS, Grenoble INP, Institut N\'{e}el, 38000 Grenoble, France}
\author{Ivan Amelio}
\affiliation{INO-CNR BEC Center and Dipartimento di Fisica, Universit\`a di Trento, 38123 Povo, Italy}
\author{Jean-Guy Rousset}
\affiliation{Institute of Experimental Physics, Faculty of Physics, University of Warsaw, Hoza 69, 02-681 Warszawa, Poland}
\affiliation{Univ. Grenoble Alpes, CNRS, Grenoble INP, Institut N\'{e}el, 38000 Grenoble, France}
\author{Jacqueline Bloch}
\affiliation{Centre de Nanosciences et de Nanotechnologies, CNRS, Universit\'{e} Paris-Sud, Universit\'{e} Paris-Saclay, C2N Marcoussis, F-91460 Marcoussis, France}
\author{Aristide Lema\^itre}
\affiliation{Centre de Nanosciences et de Nanotechnologies, CNRS, Universit\'{e} Paris-Sud, Universit\'{e} Paris-Saclay, C2N Marcoussis, F-91460 Marcoussis, France}
\author{Alberto Amo}
\affiliation{Univ. Lille, CNRS, Physique des Lasers Atomes et Mol\'ecules, F-59000 Lille, France}
\author{Anna Minguzzi}
\affiliation{Univ. Grenoble Alpes, CNRS, LPMMC, 38000 Grenoble, France}
\author{Iacopo Carusotto}
\affiliation{INO-CNR BEC Center and Dipartimento di Fisica, Universit\`a di Trento, 38123 Povo, Italy}
\author{Maxime Richard}
\affiliation{Univ. Grenoble Alpes, CNRS, Grenoble INP, Institut N\'{e}el, 38000 Grenoble, France}

\date{\today}

\maketitle

\section{Decay time of resonantly excited polaritons: evidence for coupling with a long-lived reservoir}

\begin{figure}[bt]
\includegraphics[width=1\columnwidth]{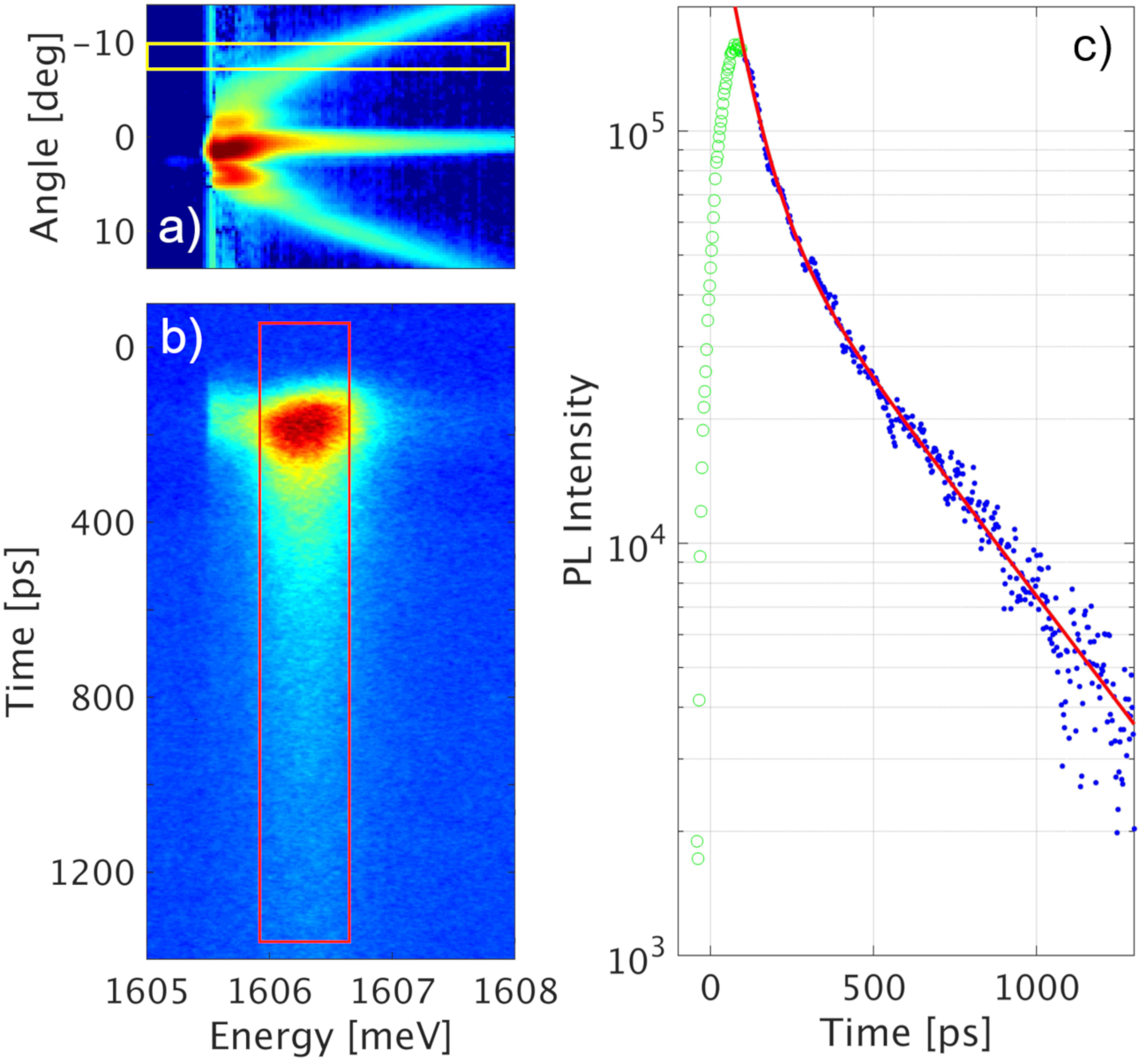}
\caption{\textbf{Time-resolved excitations PL} - a) Angle and energy resolved excitation photoluminescence obtained under weak picosecond-pulse excitation. The yellow rectangle show a cross-section at $7^\circ$ which is sent onto the streak camera. The resulting time-resolved trace is shown in b) in a time energy color plot, and the relevant cross-section $I(t)$, integrated in energy over the polariton emission linewidth (red rectangle) is shown in c) along with a fit (red line) consisting of a sum of two exponential decays with characteristic times $\tau_1=58\,$ps (instrument-limited) and $\tau_2=400\,$ps.}
\label{figS5}
\end{figure}

In order to get an independent indication that under resonant optical drive of polaritons, a reservoir of long-lived excitons is also excited in the microcavity, we performed a time-resolved photoluminescence decay measurement. The excitation strategy is similar to that used to measure the free particle dispersion relation: quasi-resonant laser excitation, red detuned central frequency with respect to the polariton ground state, weak peak intensity in order to minimize nonlinear dynamics, and small angular spread. The laser consists in picosecond pulses of $0.5\,$meV linewidth (FWHM). We checked that the laser high-energy tail has a negligible overlap with the upper polariton branch. Fig.\ref{figS5}.a shows the thus obtained time-integrated excitation photoluminescence (EPL) pattern $I_{e}(\theta,\omega)$. We select a polariton state at an angle of $7^\circ$, where contribution from the laser light is negligible (yellow rectangle in the figure) and send it into the streak-camera. The corresponding raw time-resolved trace is shown in Fig.\ref{figS5}.b and analyzed in Fig.\ref{figS5}.c. We fit it with a sum of two exponential decays, and find two characteristic timescales $\tau_{\rm res}=58\,$ps and $\tau_2=400\,$ps. \corr{The first one corresponds to the instrumental time-resolution, which is fixed by the (sharp) spectral resolution of the detection. It thus reflects a dynamics $\tau_1$ much faster than $\tau_{\rm res}$, that can be unambiguously attributed to the polariton state decay.} $\tau_2$ is too slow to be photon or polariton related, and corresponds to the typical timescale of long-lived excitons.

In order to have a clear understanding of these two timescales, we simulate the population dynamics of the coupled polariton/reservoir system by simplifying the scalar model of eq.(3) and eq.(4) into two coupled rate equations, in which we neglect the stimulation term in the back conversion mechanism considering the weak amplitude of the pump (i.e. the pump rate is much lower that the total loss rate):
\corr{\begin{eqnarray}
\partial_t n_p &=& -(\gamma_c+\gamma_{in})n_p +\gamma_{bk}n_R \label{np_decay}\\
\partial_t n_R &=& \gamma_{in}n_p-(\gamma_R^{nr}+\gamma_{bk})n_R \label{nr_decay},
\end{eqnarray}}
where $n_p$ is the condensate population, $n_R$ the long-lived excitons reservoir population, \corr{$\gamma_R^{nr}$ is the non-radiative decay rate of reservoir excitons, $\gamma_{in}$ is the conversion rate from polaritons to reservoir excitons, $\gamma_{bk}=\gamma_{in}\,{\rho_{pol}}/{\rho_{R}}$ is the reverse rate from reservoir excitons to polaritons, $\rho_{pol}$ and $\rho_{R}$ are the densities of states of polariton and reservoir states, $\gamma_R=\gamma_R^{nr}+\gamma_{bk}$}, and the other parameters are consistent with those used in the main text, and with the illustration in Fig.1.a and Fig.1.b. \corr{Note that the stimulated scattering terms proportional to the product $n_p n_R$ that stem from the bosonic enhancement factors in the quantum kinetic equations cancel out and eventually disappear from the final form of Eqs. (\ref{np_decay}-\ref{nr_decay}). Note that the $\gamma_{bk}n_R$ term in the equation for the polariton population (\ref{np_decay}) refers to the incoherent polariton population corresponding to the excitations. If needed, it could be included in the equation for the field $\psi$ (Eq.(3) of the main text) as stochastic terms.} The general solutions of this set of differential equations are of the form
$A\exp(-\Gamma_1t)+B\exp(-\Gamma_2t)$, where
\begin{equation}
\Gamma_{1,2}=\frac{1}{2}\left[\gamma_c\pm\sqrt{\gamma_c^2+4\left[\gamma_{bk}\gamma_{in}-(\gamma_{bk}+\gamma^{nr}_R)(\gamma_c+\gamma_{in})\right]} \right].
\end{equation}
Since the radiative decay of polaritons is the fastest timescale, the two $\Gamma_{1,2}$ can be simplified into
\begin{eqnarray}
\Gamma_1&=&\gamma_c \\
\Gamma_2&=&\corr{\gamma^{nr}_R+\gamma_{bk}=\gamma_R}.
\end{eqnarray}
The slow decay time component can therefore be clearly attributed to the presence of long-lived excitons from the reservoir.

\section{Determination of the dispersion relation and its confidence interval from the measurements}

\begin{figure}[bt]
\includegraphics[width=\columnwidth]{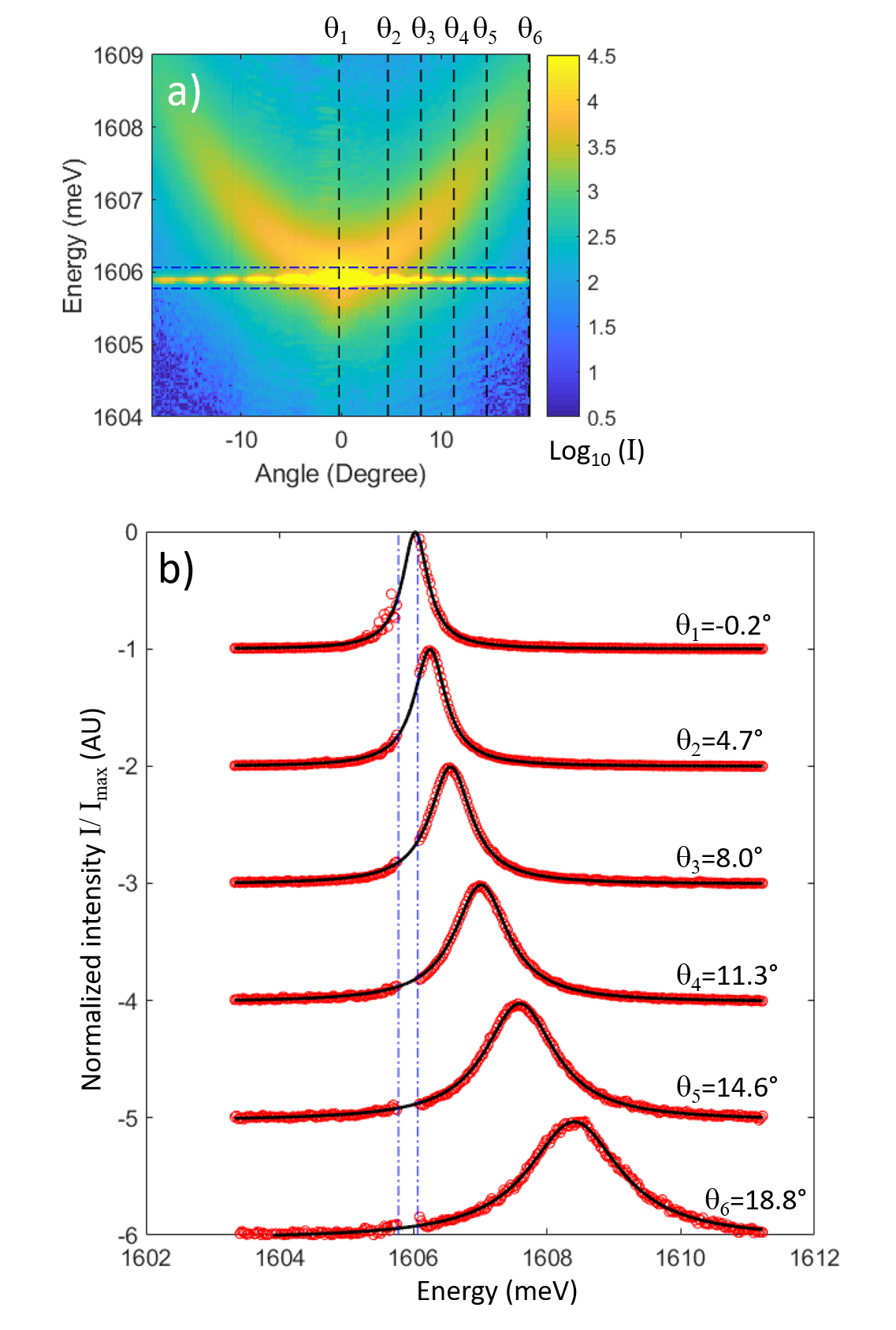}
\caption{\textbf{Experimental derivation of the dispersion relation} - a) Raw measurement of $\log_{10}[I_e(\theta,\hbar\omega)]$ of WPA point '3' (color scale). b) spectra extracted from a) at angles $\theta_i={-0.2^\circ,4.7^\circ,8.0^\circ,11.3^\circ,14.6^\circ,18.8^\circ}$ (red symbols), and Lorentzian fits (solid black line). The dash-dotted blue lines show the laser rejection interval.}
\label{figS6}
\end{figure}

We extract the experimental dispersion relations from the measured $I_e(\theta,\omega)$. A careful control of the uncertainty is a critical point in this work, as the change of shape of the dispersion relation due to the interactions is a small effect, i.e. typically much smaller than the linewidth. We thus have to make sure that the dispersion shape extracted from the measurement is not the result of some noise source or of scattered laser light. We thus carry out the following analysis :

As explained in the Methods section, for each column of CCD pixels $i$ (that contains the spectrum at a given wavevector $k_\parallel^{(i)}$), the EPL emission peak is fitted with a Lorentzian peak. from this fit, and the numerical agreement with the Lorentzian shape, we get the central frequency $\omega_i/2\pi$ of the peak as well as its $95\%$-confidence interval. The linewidth $\Delta \omega_i$ and its $95\%$-confidence interval are obtained as well from this procedure. The results for WPA and WPB are shown in the main text in Figs.3, where the line thickness of the experimental plots shows the $95\%$ confidence interval.

Such an analysis is illustrated in Fig.\ref{figS6} for the data from the working point WPA.3 defined in the main text. Fig.\ref{figS6}.a show the raw measurement of $I_e(\theta,\omega)$ in color log scale. Six spectra extracted from six different angles ($\theta_1$ to $\theta_6$) are shown in Fig.\ref{figS6}.b in red. The filter rejecting the laser is shown as the interval between the blue dash-dotted lines. The Lorentzian fit is shown as a solid black line for each spectra. The quantitative results of this fit are summarized in the table below
\\
\\
\begin{tabular}{c||c|c|c|c|}
i & $\theta_i$ [Deg] & $R^2$ & $\hbar\omega_0$ [meV] & $\hbar\Delta\omega_0$ [meV] \\
\hline	
1 & -0.2$^\circ$    & 1.31\% & 1606.027        & 0.011 \\
2 &  4.7$^\circ$    & 0.07\% & 1606.246        & 0.003 \\
3 &  8.0$^\circ$    & 0.09\% & 1606.555        & 0.004 \\
4 & 11.3$^\circ$    & 0.10\% & 1607.012        & 0.005 \\
5 & 14.6$^\circ$    & 0.11\% & 1607.591        & 0.007 \\
6 & 18.8$^\circ$    & 0.22\% & 1608.402        & 0.012
\label{tab2}
\end{tabular}
\\
\\
\\
\begin{tabular}{c||c|c|}
i & $\hbar\gamma$ [meV] & $\hbar\Delta\gamma$ [meV] \\
\hline	
 1 & 0.466 & 0.014 \\
 2 &0.573 & 0.004  \\
 3 & 0.748 & 0.007 \\
 4 & 0.974 & 0.009 \\
 5 & 1.274 & 0.012 \\
 6 & 1.676 & 0.025
 \label{tab3}
\end{tabular}
\\
\\
\\
The uncertainties $\hbar\Delta\omega_0$ and $\hbar\Delta\gamma$ are the 95\% confidence interval (sometime referred to as the $2\sigma$ confidence interval) of $\hbar\omega_0$, the central energy of the peak, and $\hbar\gamma$ its full-width at half-maximum, as obtained from the fitting algorithm.

\section{Excitation mechanisms of the condensate by thermal phonons}

\begin{figure}[bt]
\includegraphics[width=0.65\columnwidth]{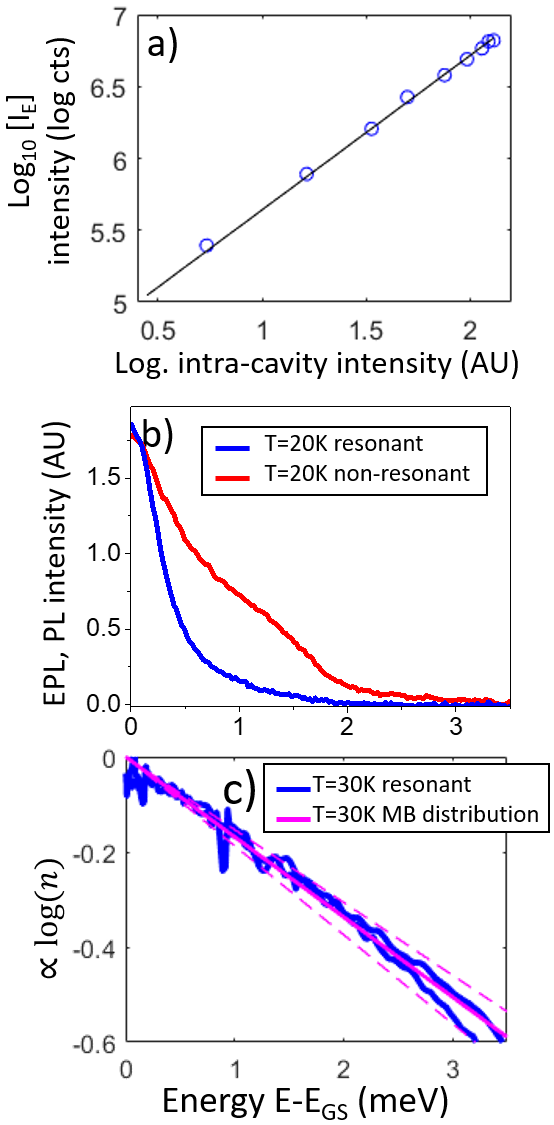}
\caption{\textbf{Linearity of the transfer mechanism and polaritonic population distribution} - a) Blueshift-corrected polariton excitation density versus cavity-coupled laser intensity. b) Angle-integrated spectra of polaritons photoluminescence under resonant (blue) and non-resonant laser excitation (red). c) Polariton occupancy $n$ (within a constant factor) versus energy in the free particle regime (blue). Calculated Maxwell-Boltzmann distribution at $T=30\,$K (solid magenta), $T=33\,$K and $T=27\,$K (dashed magenta).}
\label{figS2}
\end{figure}

\begin{figure*}[hbt]
\includegraphics[width=0.7\textwidth]{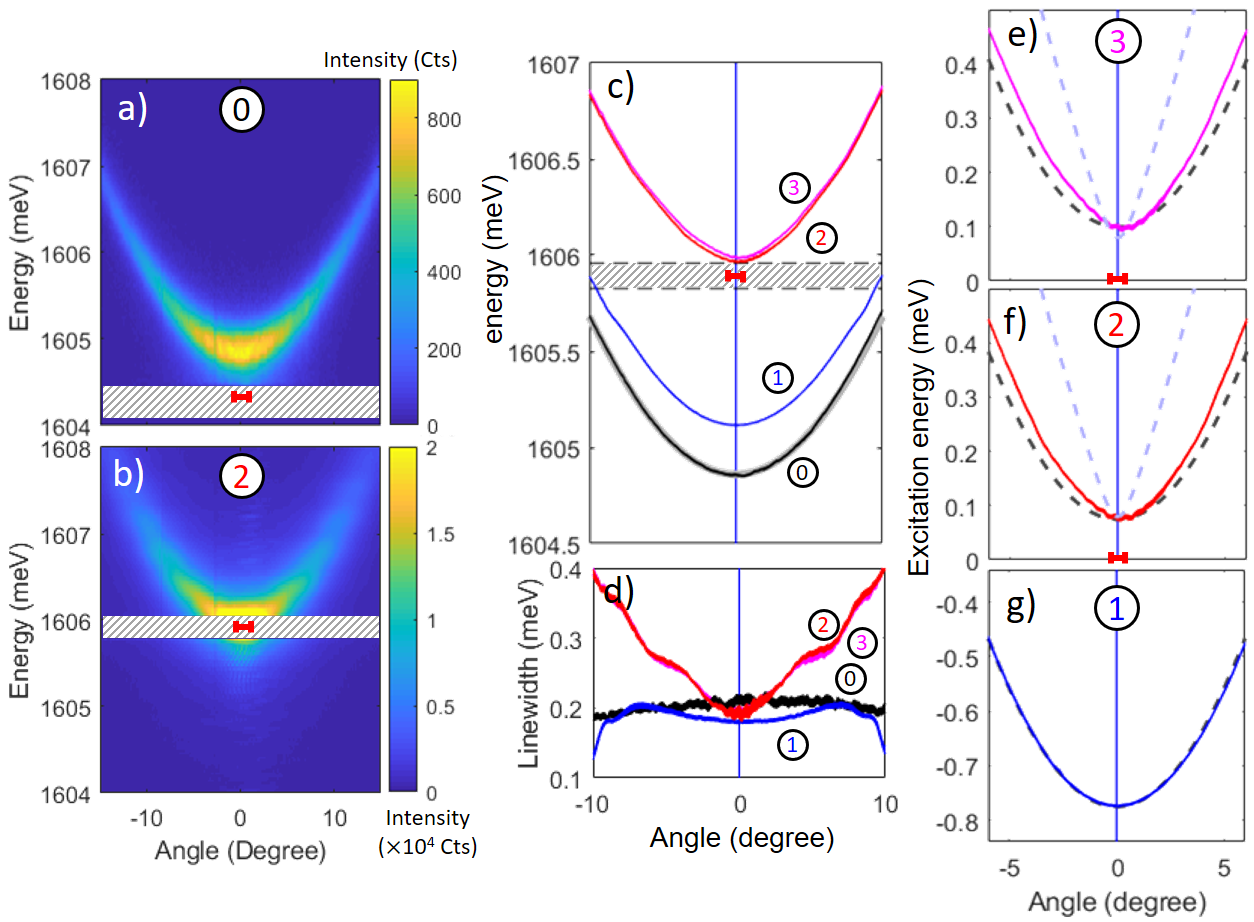}
\caption{\textbf{Excitation dispersion relation of Working point C} -
Angle and energy resolved cross-polarized EPL measurements $I_e(\theta_p,\hbar\omega)$ in the free particle regime (a), and in the upper branch of $I_0(P)$ (b). The intensity is color-coded on a linear scale. The laser energy and angular spread is shown as a red segment. (c) Measured dispersion relations $\hbar\omega(\theta)$ as obtained from the numerical analysis (solid colored lines). The line thickness show the $95\%$-confidence interval of the experimentally determined dispersion relations. Four states of the fluids are shown labelled from 0 to 3: the free particle dispersion ('0'), a lower branch state of $I_0(P)$ ('1'), and two upper branch states ('2' and '3'). Panel (c) show the dispersions relations measured on absolute energy scale, while (e-g) show the dispersion relations '3' to '1' respectively measured from the condensate energy. The hatched rectangle in (a-c) show the spectral range rejected by the notch filter. (d) Measured excitation linewidth versus angle of states '0' to '3'. Panels (e-g) include two calculated limiting cases: the vanishing condensate fraction regime (black dashed line), and the regime of vanishing reservoir fraction (blue dashed line).}
\label{figS4}
\end{figure*}

In this section, we present in details the arguments that support the fact that thermal phonons are the likeliest origin of spontaneously created excitations on top of the condensate observed in this work. We have shown in another work, involving microcavities made up of another materials, and at higher temperatures, that there are several possible mechanisms that can create such excitations \cite{klembt_2015}:
\begin{enumerate}

\item Auger scattering of two polaritons from the condensate: the recombination of a polariton excites a hot electron-hole pair, that relax into the long-lived reservoir of free excitons (at $\omega \geq \omega_X$, and momentum beyond the light cone), and end up exciting fluctuation on top of the condensate. We see that this case involves a reservoir of free excitons that can interact with the polariton condensate as described in our work.

\item Inelastic scattering by thermal optical phonons: This mechanism takes a polariton from the condensate and create a free exciton in the reservoir described above. This inelastic scattering is possible in any microcavity as long as the polariton condensate energy is separated from the large momentum excitons reservoir by less than the energy of an optical phonons. This condition is largely met in GaAs-based microcavity. Note however that at T=$30K$, the temperature of our experiment, the occupancy of the optical phonons population is very small: it amounts to $\exp(-\hbar\omega_{LO}/k_BT) \simeq 1\times 10^{-6}$, where $\hbar\omega_{LO}=36.5\,$meV is the longitudinal optical phonon energy in GaAs.

\item Inelastic scattering by thermal acoustic phonons: Such a mechanism allows out-scattering of polaritons towards any state within the frequency window $[0,\omega_{LO}]$. Considering the frequency dependence of this thermal population, scattering towards states of low frequency with respect to the condensate are largely favoured. These states are (i) excitations of the polariton condensate, and (ii) localized states of excitons caused by imperfections in the quantum well, that can lie several meV below the free exciton energy, i.e. close or at resonance with the polariton condensate. These imperfections typically consist in thickness and/or alloy fluctuations \cite{borri_2000,savona_2007}.

\end{enumerate}

In scenario (1), we expect that the relation between the excitation power entering the cavity $I_{\rm in}$ and the polariton condensate population $I_0$ is sublinear. We thus checked this possibility by exciting polaritons resonantly in the redshifted configuration, i.e. $\Delta=\hbar\omega_{l}-\hbar\omega^0=-0.48\,meV$, and by measuring the angle-integrated EPL intensity versus the excitation power. What matters is the relation between the laser intensity effectively entering the cavity spacer $I_E$, that depends both on the outside excitation power $I_{\rm out}$ and on the blueshifting polariton mode $E_{\rm LP}(I_{\rm out})$. This relation can be accurately modeled in the weak to moderate excitation power by assuming a lorentzian lineshape of fixed linewidth for the polariton mode, such that
\begin{equation}
I_{\rm in}\propto \left[\left(E_{\rm las}-E_{\rm LP}(I_{\rm out})\right)^2+(\hbar\gamma_c)^2\right]^{-1},
\end{equation}
where both $E_{\rm LP}(I_{\rm out})$ and $\hbar\gamma_c$ are direct experimental observables. Assuming that the angle-integrated EPL intensity is proportional to the condensate density, we can thus measure the relation between $I_0$ and $I_{\rm in}$. The result is shown in Fig.\ref{figS2}.a: it turns out that over 1.5 orders of magnitude, this relation is linear (or slightly superlinear), with a fitted exponent of $1.07\pm0.03$. Note however that this particular argument should be taken with caution, it is based on the assumption that the magnitude of the birefringence that couples the co- and cross-polarized condensate does not depend on the nonlinearity, which might not necessarily be true.

Another, and perhaps even stronger argument which is inconsistent with scenario (1), as well as with scenario (2) is the following: both scenarios involve an intermediate reservoir of long-lived free excitons beyond the light cone. In these scenario, this reservoir relaxes into condensate excitations by two-body scattering or by emission of phonons. A way to check this possibility is to compare the condensate excitation spectrum $I_{R}(\hbar\omega)$ obtained under resonant excitation, with the one measured under non-resonant excitation, when the laser is tuned far above ($100\,$meV) the free excitons states. In the latter case, the relaxation mechanism involves the same intermediate reservoir as in scenario (1) and (2) such that the thus obtained excitation spectrum $I_{NR}(\hbar\omega)$ should be similar in shape to $I_{e}(\hbar\omega)$. The comparison is shown in Fig.\ref{figS2}.b at the same working point and temperature $T=20\,$K: It turns out that these spectra are quite different, the one obtained under non-resonant excitation being significantly 'hotter'.

Scenario (3) is the most likely one in our experiment. We can support it further by checking that the condensate fluctuations are indeed excited by the thermal bath of acoustic phonons by measuring the polariton state occupancy $n(\hbar\omega)$ in the free particle regime (vanishing two body interactions), and compare it with a thermal distribution. $n(\hbar\omega)$ can be determined experimentally using the density of state $\rho(\hbar\omega)$ effectively contributing in the measurement, and $I_{e}(\hbar\omega)$ which is proportional to the fluctuation population times their linewidth. The measurement has been performed at $T=30\,$K and the result is shown in log scale in Fig.\ref{figS2}.c. The excitations occupancy $n(\hbar\omega)$ obtained in this way agrees very well with a Maxwell-Boltzmann distribution at $T=30\,$K. The distributions for $T=27\,$K and $T=33\,$K are shown as dashed lines. This result is a strong argument in favour of scenario (3).

\section{Working point C}

In order to support the generality of our observations, we have performed a full characterization of the polariton superfluid excitations at several other working points. Fig.\ref{figS4} shows the characterization at working point C, in which the condensate is driven at normal incidence $\theta=0^\circ$. Fig.\ref{figS4}.a shows the raw EPL measurements in the red-shifted weak excitation regime, that we use to determine the reference free-particle dispersion, and Fig.\ref{figS4}.b shows the raw EPL measurement in a switched-up state. The hysteresis curve $I_0(P)$ is shown in Fig.\ref{figS1}.a. The analyzed dispersion relations are shown all together in and Fig.\ref{figS4}.c and individually in Fig.\ref{figS4}.e-g. In the latter case, the limiting cases theory (i) and (ii) discussed in the main text are plotted for comparison as black and blue dashed line respectively. The extracted linewidth analysis is plotted as well in Fig.\ref{figS4}.d. As explained in the next section, the flat behaviour of $\hbar\gamma$ vs angle for the free-particle dispersion and the switched down states results from the absence of switching front, and of a spatial filter in the detection. The experimental parameters of WPC are $\delta=+0.7meV$, $T=30$K, free particle linewidth at $k=0$ $\hbar\gamma_c=0.4$meV, and $\Delta=+1.04$meV.

\section{characteristics of the condensate in real space}

\subsubsection{Spatial structure of the condensate}
\begin{figure}[bt]
\includegraphics[width=\columnwidth]{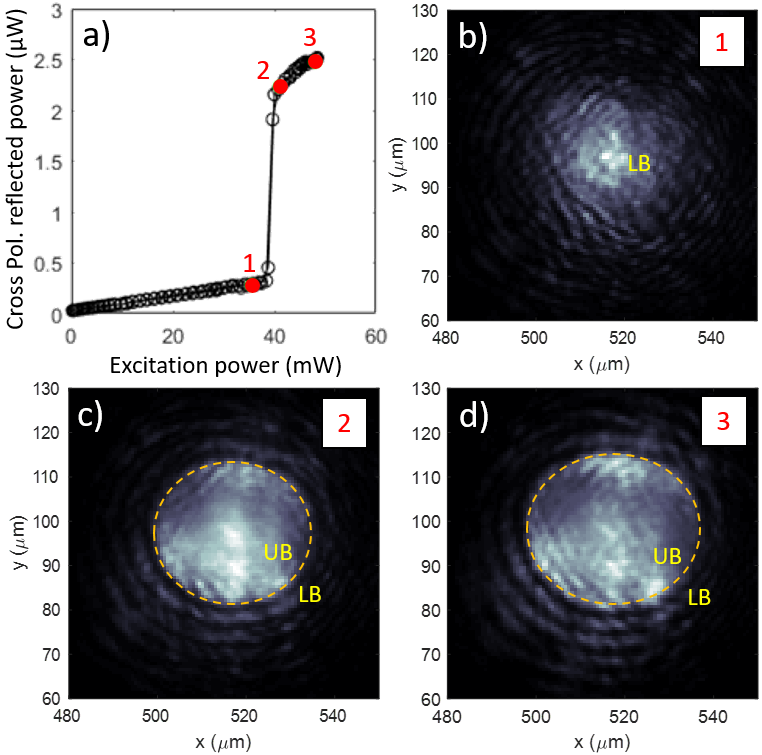}
\caption{\textbf{condensate spatial structure} - (a) Measured hysteresis curve $I_0(P)$ of the cross-polarized reflectivity for a working point with parameters identical to WPA. The cross-polarized condensate density is shown in real space for the states '1' (b), '2' (c) and '3' (d) marked on the hysteresis curve. 'LB' ('UB') stands for lower (upper) branch of $I_0(P)$. The dashed orange line shows the circular switching front separating the 'LB' and the 'UB' areas.}
\label{figS1}
\end{figure}

Owing to the Gaussian shape of the excitation spot the upper branch of the hysteresis is located within a round area at the center of the excitation spot, separated by a sharp switching front of diameter $D^{(2)}$ from the outer ring of polaritons, that are in a switched-down state. This is shown in Fig.\ref{figS1} where the measured cross-polarized condensate at WPC is shown in real space for three different regimes: in the state labeled '1' in Fig.\ref{figS1}.b the whole fluid is in the lower branch of the hysteresis (which is shown in  Fig.\ref{figS1}.a), and thus exhibits a gaussian intensity distribution. In Fig.\ref{figS1}.c the central part of the fluid labeled 'UB' has switched to the upper branch, while the outer rim labeled 'LB' remains in the lower branch. The switching front separating the two is shown as a dashed orange line. At higher intensity (state '3' in Fig.\ref{figS1}.d), the diameter of the switching front slightly increases, i.e. moves further towards the edges of the fluid.

\subsubsection{Influence on the measured linewidth $\hbar\gamma(\theta)$ in the high-density regime}
\begin{figure}[bt]
\includegraphics[width=0.55\columnwidth]{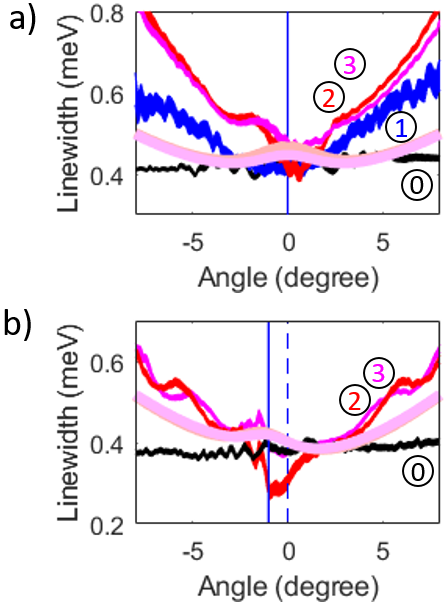}
\caption{\textbf{Linewidth versus momentum} - Measured excitations linewidth $\hbar\gamma(\theta)$  vs Angle, as obtained from the numerical analysis of the EPL (narrow solid lines) for WPA (a) and WPB (b). The full vectorial theory is shown as the thick solid lines. For the measurement data, the line thickness show the $95\%$-confidence interval. The same labelling convention is used as in Fig.3 of the main text. The solid vertical line show the laser excitation incidence angle, while the dashed one shows the zero incidence reference ($\theta_p=0^\circ$)}
\label{figS7}
\end{figure}

The analysis of the raw measurement $I_e(\theta,\omega)$ also gives us access to the excitations spectral linewidth $\hbar\gamma$. The results are shown in Fig.\ref{figS7} for both WPA (a) and WPB (b) together with the results of the full theory calculation. In the low excitation regime, (solid black line labelled '0' in WPA), the measured linewidth is essentially angle-independent. This is expected as over such a small angular range, the excitonic and photonic fraction hardly change vs $\theta$: i.e. the kinetic energy increase as compared to the Rabi splitting is small ($1\,$meV increase between $0^\circ$ and $10^\circ$). For the measurements of $\hbar\gamma(\theta)$ in states '1' (switched-down state), '2', and '3' (switched-up states), we have inserted a circular aperture of diameter $D^{(1)}$ in the detection path that filters out the outer area of the excitation spot.

For the measurement of $\hbar\gamma(\theta)$ in states '2' and '3', both the finite size of the 'UB' area in real space, and the use of a circular aperture filter have a significant influence: in both cases, the excitations of the condensate have a finite transit time either throughout the circular aperture of the detection or within the switched up area, depending on which is the smallest. This transit time is fixed by the excitation group velocity $\tau(k)=D^{(1,2)}/v_g(\theta)$, which is a decreasing function of $\theta$, regardless of the detailed nature of the excitation. For the switched down state shown in Fig.\ref{figS1}.a, there is no switching front such that the circular aperture of diameter $D^{(1)}=40\,\mu$m actually fixes the increasing rate of $\bar\gamma$ versus $\theta$. This is the case of state '1' in Fig.3\ref{figS7}.a in the main text. For states labeled '2' and '3' in Fig.\ref{figS7}.a and Fig.\ref{figS7}.b, the switching front diameters $D^{(2)}$ are comparable and close to $ D^{(1)}=40\,\mu$m. As a result, the increase of $\hbar\gamma$ versus $\theta$ reflects the presence of the switching front, and can thus be considered as a characteristic feature of a locally switched-up polariton fluid driven by a Gaussian spot.

While our full theory is translational invariant, we can still mimic this effect with it by adding a numerical spatial filter of diameter $D^{(2)}$ to the results. The result is shown is Fig.\ref{figS7} as the light solid lines. We see that while the increasing linewidth versus $|\theta|$ is indeed reproduced, it is underestimated. Other sources of nonlinear broadening are likely to contribute for
increasing excitations energy, as the reservoir becomes highly populated (not accounted for in the present model).

\subsubsection{Influence of the Gaussian spot on the dispersion relation of the excitations $\omega_e(\theta)$}

\begin{figure}[bt]
\includegraphics[width=0.8\columnwidth]{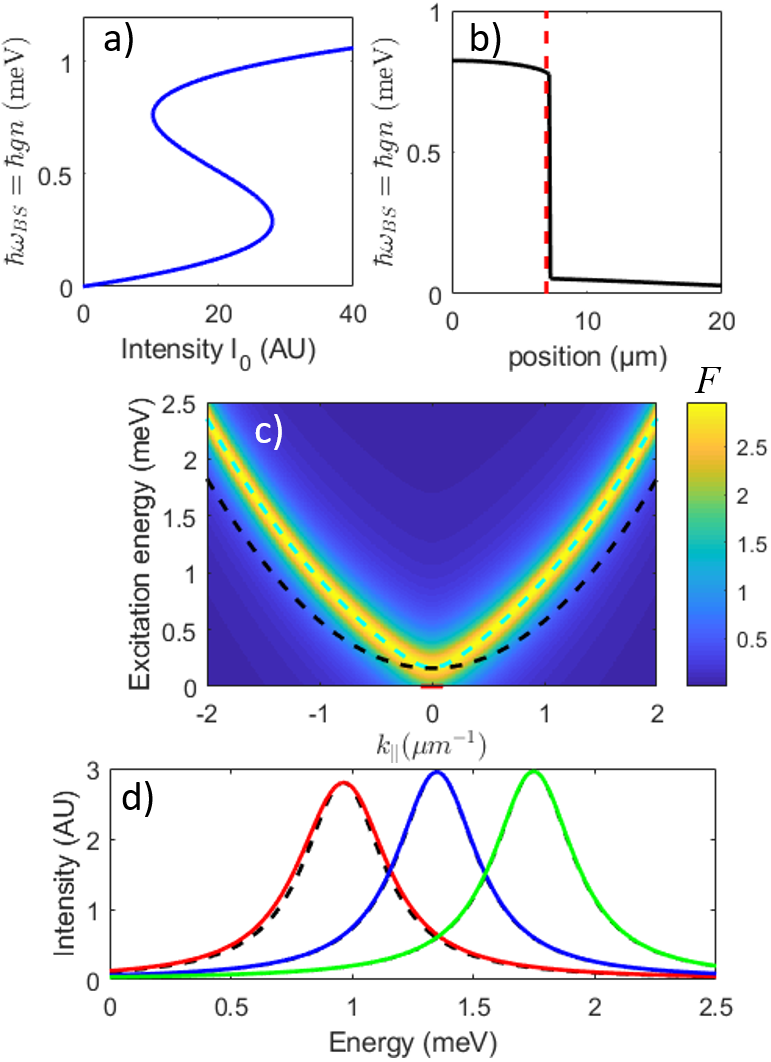}
\caption{\corr{\textbf{influence of the Gaussian spot on the dispersion relation} - a) calculated $n(I_0)$ for $\hbar\Delta=-0.79\,$meV, and $\hbar\gamma=0.4\,$meV from the homogeneous steady-state GGPE. b) calculated Profile of the polariton density $n(r)$ in the LDA, for a Gaussian spot of size $\sigma_r=25\,\mu$m and a peak intensity $I_0$ close enough to the switch down point in order to match the experimental blueshift (of WPA.2) $\hbar\omega_{BS}=gn=0.85\,$ meV. The dashed red line shows the spatial filter edge used in the calculation. c) Calculated inhomogeneous spectral function $F(k_\parallel,\omega)$ in the LDA. The cyan dashed line is the theoretical excitations dispersion relation of the homogeneous pure condensate. The black dashed line is the homogeneous rigid blueshift $\omega_{RB}(k_\parallel)$. d) Cross-sections of $F(k_\parallel=k_j,\omega)$ for $k_j={0,0.6,1}\,\mu$m$^{-1}$ in red blue and green respectively. The dashed line show the homogeneous lineshape $\mathcal{L}(\omega,\omega_0,\gamma)$ for comparison.}}
\label{fig_R2}
\end{figure}


The density profiles shown  in Fig.\ref{figS1}.c,d display two types of inhomogeneities: an overall modulation due to the pump Gaussian intensity profile, and random fluctuations due to in-plane disorder. We discuss here how the Gaussian spot may influence the dispersion relation, and in the next section the influence of disorder.

This problem is difficult to tackle in the framework of the full theory: the specific shape of the cloud strongly depends on the bistability features and, in turn, on the details of the switch-off jump at the edge of the cloud. However, we can take the simplifying assumptions of the local density approximation (LDA), which is reasonable, if not exact, considering the large size of the Gaussian spot ($\sigma_r=25\,\mu$m radius).

We thus assume a pure, scalar polariton condensate for which the excitations dispersion relation $\omega_{Bog}(k_\parallel,\omega,n)$ is given by Eq.(2) of the main text. The pump spot has a bidimensionnal Gaussian intensity profile: $I_P (r)=I_0/(\pi\sigma_r^2)\exp\left[-(r^2/\sigma_r^2)\right]$. The local polariton density $n(r)$ is derived within the LDA from the steady-state GGPE equation $I_P(n)=n\left[(\Delta+gn)^2+(\gamma/2)^2 \right]$.

Using the experimental parameters $\hbar\Delta=-0.79\,$meV, and $\hbar\gamma=0.4\,$meV (WPA), we obtain the hysteretic plot $I_P(n)$ shown in Fig.\ref{fig_R2}.a, and the density distribution $n(r)$ (Fig.\ref{fig_R2}.b). This density profile is in qualitative agreement with the measured one shown in Fig.\ref{figS1}.c,d in terms of its 'flatness' in the switched-up area (excluding the fluctuations due to the weak disorder). Assuming a Lorentzian lineshape, the inhomogeneous spectral function $F(k,\omega)$ is obtained in the LDA approximation as
\begin{equation}
F(k_\parallel,\omega)=\iint_f {\rm d}x{\rm d}y\, n(r)\mathcal{L}\left(\omega,\omega_{Bog} \left[k_\parallel,\omega,n(r)\right],\gamma\right),
\end{equation}
where $n(r){\rm d}x{\rm d}y$ is proportional to the local intensity ${\rm d}I(x,y)$, $\mathcal{L}(\omega,\omega_0,\gamma)$ is a Lorentzian lineshape centered at $\omega_0$ and of linewidth (FWHM) $\gamma$, and $f$ is the spatial filter (circular aperture) which is centered on the switched-up area, with a smaller diameter in order to reject the switching front separating the two regions, like in the experiment (red dashed line in Fig.\ref{fig_R2}.b). The calculated spectral function is shown in Fig.\ref{fig_R2}.c in colorscale. Like in the manuscript, we can compare it with the two limiting cases: \emph{homogeneous} pure polaritons condensate dispersion relation (cyan dashed line), and rigidly blueshifted dispersion relation (black dashed line), both with a blueshift equal to the experimental one ($\hbar\omega_{BS}=0.85\,$meV, WPA.2). We see very clearly that in spite of the inhomogeneity from the Gaussian spot, the fit with the pure condensate limit is nearly perfect. This is in contradiction with the measurement, in which the speed of sound is found twice too low as compared to the homogeneous pure condensate limit. This is far too large a deviation, even in the local density approximation.

Note that the LDA being an approximation, we tested parameters slightly different from the experimental ones in terms of linewidth and laser detuning $\Delta$, for which the switched up density distribution is less flat. The deviation between the homogeneous dispersion relation and the inhomogeneous spectral function remained also largely negligible.

This analysis provides another relevant insight. In spite of using a Gaussian spot for the excitations, we see that the hysteretic behavior offers a 'flattening' of the polaritons density, which limits the inhomogeneous broadening of the excitations. This is also true in the case where a reservoir is involved.  In essence, the nonlinearity acts as a spatial 'tophat' filter for the Gaussian pump spot, a strategy often used in optical spectroscopy to mimic a quasi-homogeneous spot intensity distribution. The positive influence of this flattening effect is visible (within the LDA) in Fig.\ref{fig_R2}.d, where inhomogeneous spectra $F(k_\parallel=k_j,\omega)$ for three values of $k_j$  are plotted (color lines) and compared with the homogeneous lineshape $\mathcal{L}(\omega,\omega_0,\gamma)$ (black dashed line). In spite of the Gaussian excitation, we see a vanishing deviation between the two. This is in agreement with the experimentally measured linewidth shown in Fig.\ref{figS7}. Indeed, if we look at the excitation spectrum linewidth at $\theta=0$, for which the effect of the spatial filter is minimum (due to vanishing group velocity), we see that in spite of a total blueshift of 0.85\,meV (Fig.\ref{figS7}.a red line, $\theta=0^\circ$), the measured linewidth of 0.4\,meV remains unchanged as compared to that in the non-interacting regime (Fig.\ref{figS7}.a black line, $\theta=0^\circ$), within the experimental uncertainty. This is a solid indication that, thanks to the nonlinear 'flattening' of the density, the Gaussian spot shape has a negligible influence on the excitations lineshape and on the dispersion relation shape.

\subsubsection{Influence of disorder on the dispersion relation of the excitations $\omega_e(\theta)$}

\begin{figure}[bt]
\includegraphics[width=0.8\columnwidth]{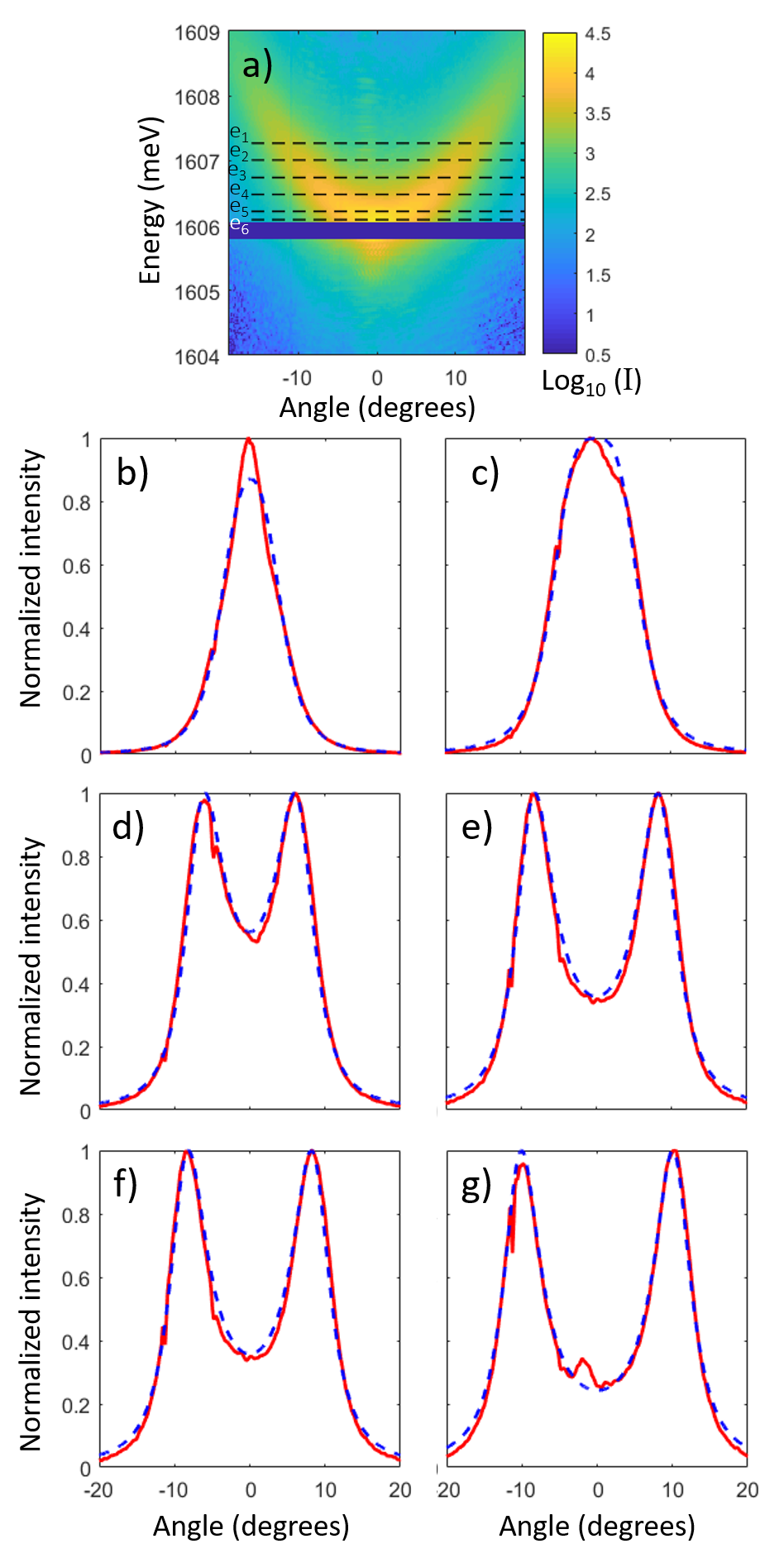}
\caption{\corr{\textbf{Angular cross-sections} - a) Raw measurement of $\log_{10}[I_e(\theta,\hbar\omega)]$ of WPA point '3' (color scale). Experimental (red line) and theoretical (dashed blue line) horizontal cross sections $I_e(\theta,e_i)$ are shown for $e_6=1606.09\,$meV (b), $e_5=1606.22\,$meV (c), $e_4=1606.48\,$meV (d), $e_3=1606.74\,$meV (e), $e_2=1607.01\,$meV (f), qnd $e_1=1607.27\,$meV (g). The small peak visible in panels b,c and g around $\theta=-2^\circ$ is a residue from the pump laser.}}
\label{figS8}
\end{figure}

The darker and brighter patches observed in Fig.\ref{figS1}.c,d are attributable to a local disorder of weak amplitude (as compared to the linewidth) and correlation length in the $3-5\,\mu$m (cf. Fig.2.b in the main text). Again, they are of unknown relative amplitude with respect to the total condensate density. Owing to the relatively small correlation length, the LDA approximation is probably not good enough to analyze their influence on the excitation dispersion relation. However, a disruptive disorder should result in (i) a significant broadening and a possible asymmetry of the measured spectra $I(\theta=\theta_j,\omega)$, and (ii) the signatures of localization: spectrally narrow states, broad in momentum space, and flat in dispersion.

We have looked for both signatures in the experimental data and found none of them. Measured spectra $I(\theta=\theta_j,\omega)$ are shown in detail in Fig.\ref{figS6}.b, and exhibit no sharp peaks indicative of localization within the experimental uncertainty. We have also looked at cross-sections at fixed energies $I(\theta,\omega=\omega_j)$ (red lines in Fig.\ref{figS8}), and compare them with the spectral function of the free-polariton dispersion $F_p(\theta,\omega=\omega_j)$ (dashed blue lines in Fig.\ref{figS8}). While this is obviously an approximate model, it is sufficient to check whether some anomalous broadening in momentum space are present in the experimental peaks. The linewidths used in this simple comparison are extrapolated from the table in section II of the SI. Very small corrections to the cross-section energy are applied in order to match angular-peaks from the free-polariton model and the experiment. It is clear from this comparison that we do not see any momentum broadening or signatures of localization.

\section{Statistical analysis of the condensate fraction and renormalized speed of sound}

\begin{figure*}[bt]
\includegraphics[width=0.65\textwidth]{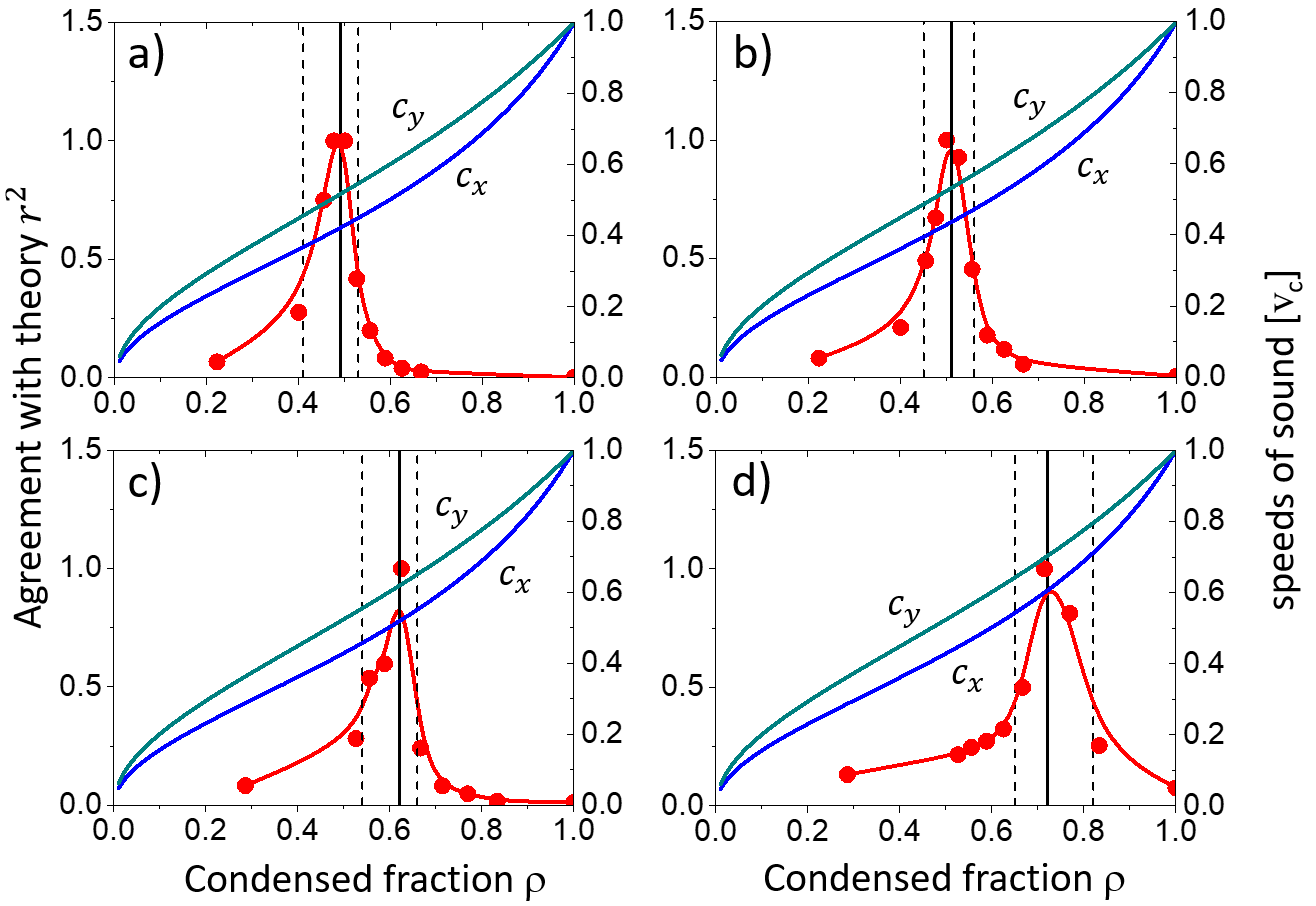}
\caption{\textbf{Statistical analysis of the condensate fraction by quantitative comparison between experiment and theory} - Agreement between the measured dispersion relation and the full vectorial theory, quantified as $r^2(\rho)=$min$(R^2)/R^2(\rho)$ (red symbols), versus the condensate fraction $\rho=n_c/(n_r+n_c)$ for WPA.2 (a),WPA.3 (b), WPB.2 (c) and WPB.3 (d). The red solid line is a guide for the eye. The right axis show the calculated speed of sounds of the two cross polarized excitations $c_x/v_c$  (blue line) and $c_y/v_c$ (blue green line) versus $\rho$, in the limit where $n_y=0$ (see text).}
\label{figS3}
\end{figure*}

The comparisons between the theoretical dispersion relations and the measured ones, are shown quantitatively in Fig.\ref{figS3} using the normalized deviation $R^2(\rho)=\sum_i [\omega_{\rm exp}(k_i)-\omega_{\rm th}(\rho,k_i)]^2/\sum_i \omega_{\rm exp}(k_i)^2$. Fig.\ref{figS3} plots $r^2(\rho)={\rm min}(R^2)/R^2(\rho)$ where
\begin{equation}
\rho=\frac{n_c}{n_R+n_c}=\frac{\gamma_R}{\gamma_{in}+\gamma_R},
\end{equation}
is the condensate fraction, $n_c=n_x+n_y=|\psi_x^s|^2+|\psi_y^s|^2$, $\gamma_R=1.6\,\mu$eV as extracted from the decay time found in Fig.\ref{figS5}, and $\gamma_{in}$ is the actual fit free parameter. The closer to $r^2$ is to 1, the better the agreement. The full width at half maximum of $r^2(\rho)$ provides an estimate of its 1$\sigma$-confidence interval. The line connecting the dots in Fig.\ref{figS3} is a guide for the eye. The thus determined confidence intervals has been used in Fig.3 of the main text, where the theoretical dispersion plots is obtained by plotting the two dispersion relations calculated at the two confidence interval boundaries, and by coloring the area that they delimit.

This analysis also gives an estimate of the two speeds of sound that characterizes the two cross-polarized Bogoliubov branches at the sonic point of the hysteresis. In the simplifying assumption that $n_y=0$, i.e $\rho=n_x/(n_x+n_R)$, they have the analytical expression given in eq.(\ref{cx}) and eq.(\ref{cy}). The results are shown in Fig.\ref{figS3} (right axis), where the speeds of sound $c_x$ and $c_y$ normalized to the critical velocity $v_c=\sqrt{\hbar\omega_{BS}/m}$ are plotted alongside $r^2(\rho)$.

The results of this whole analysis are summarized in the following table.
\\
\\
\corr{
\begin{tabular}{c||c|c||c||c}
WP & $\rho$ & CI($\rho$) & $c_x/v_c$ & $c_y/v_c$ \\
\hline	
A.2 & 47\%  & [40\%;53\%]  & 43\%   & 48\%\\
A.3 & 45\%  & [38\%;50\%]  & 42\%   & 46\% \\
B.2 & 78\%  & [70\%;85\%]  & 61\%   & 68\% \\
B.3 & 68\%  & [60\%;73\%]  & 52\%   & 57\%
\label{tab1}
\end{tabular}
}
\\
\\
where 'CI' stands for 'confidence interval'.

\section{Estimate of the polariton-polariton interaction $g_T$}

A key advantage of this work is that we can extract the relative contribution of polariton-polariton interaction $\hbar\bar{g}n$ to the total blueshifts $\hbar\omega_{BS}=\hbar\bar{g}n+\hbar g_Rn_R$. Thus for instance in WPA point 2, out of $\hbar\omega_{BS}=0.85\,$meV, $\hbar\bar{g}n=0.15\,$meV come from polariton-polariton interaction.

However, in order to derive an actual value of the polariton-polariton interaction constant $g_T$ (where $2\bar{g}=g_T+g_S$), we need additional information such as an absolute estimate of the polariton density $n$, or of the pump term $|F|^2$. Since we have a solid measurement of the latter in our $I(P)$ measurement, we can use it to infer $g_T$.

Assuming the scalar limit of our model, in the homogeneous and steady state regime, we get that
\begin{equation}
\left[(-\hbar\Delta+\hbar g_{\rm eff}n)^2+\frac{\hbar^2\bar{\gamma}^2}{4}\right]n=|F|^2,
\end{equation}
where $g_{\rm eff} = \bar{g} + g_R\gamma_{in}/\gamma_R$ and $\bar{\gamma}= (\gamma_c + \gamma_{in})$. The input-output theory of polaritons \cite{carusotto_2013} provide a relation between the intracavity polaritonic field generation rate $|F|^2$ and the external pump of power $P_in$:
\begin{equation}
|F|^2=\frac{\hbar\gamma}{2}|C|^2f_p(r_a)\frac{P_{in}}{\hbar\omega_l}\frac{10^6\hbar^2}{e^2},
\end{equation}
where $|F|^2$ is expressed in [meV.$\mu$m$^2$] units, $|C|^2$ is the photonic fraction, $P_{in}$ is the laser power in [W], $f_p(r)$ is a normalized Gaussian distribution describing the spot in $\mu$m$^{-2}$ units, and $r_a$ is an average radial position within the spot such that the spatially averaged blueshift $\langle \hbar g_{\rm eff}n\rangle$ is that found in the experiment. In introducing $f_p(r)$ and neglecting the kinetic term in the equation of motion above, we have made the implicit assumption of the slowly varying envelope. In order to determine $r_a$ we would need to know the distribution $n(\mathbf{r})$. We do not know it, but we have a measurement of the radius $r_t$ of the UB in real space shown in Fig.\ref{figS1}.c: $r_t=17.5\,\mu$m.

We thus consider two extreme situations: $r_a=0$ and $r_a=r_t$ as forming a first contribution to the error bar of $g_T$. We then fit the theoretical hysteresis $\hbar\omega_{BS}(P_{in})$ derived from the above equations such that (i) the jump-up occurs at the experimental one: $P_{ju}=30.1\,$mW as shown in $I(P)$ in Fig.2.a, and (ii) the total blueshift $\hbar g_{\rm eff}n$ matches the experimental one $\hbar\omega_{BS}=0.85\,$meV of point 2 at $r_a=0$ or $r_a=r_t$.

In doing so, we find $g_T=8\pm2\mu$eV$\mu$m$^2$ in which the error bar is actually rather set by the larger uncertainty on $\gamma_{in}/\gamma_R$ than on $r_a$. This value is well in-line with the literature, with the advantage of ruling out for sure the reservoir contribution. We insist however that this method is approximate, as it doesn't take into account the spatial shape of the spot rigorously, which would require a much more involved modelization.

\section{Full vectorial Theory}

\subsection{Theoretical model}

In order to model the experiment, we use a generalized Gross-Pitaevskii theory for the two polariton condensate wavefunctions  $\psi_\sigma({\mathbf r})$, corresponding to the two linear polarizations with $\sigma=x,y$ (pump-probe basis), coupled to a dark-exciton reservoir with density $n_R$:
\begin{widetext}
\begin{equation}
i\partial_t \psi_x =
\left[ \omega_{LP} (\hat{\mathbf{k}}) - \frac{\alpha}{2} \cos(2\Theta) + \frac{g_T+g_S}{2} |\psi_x|^2 + g_T |\psi_y|^2 + g_R n_R  -i \frac{\gamma_c + \gamma_{in}}{2} \right] \psi_x - \frac{\alpha}{2} \sin(2\Theta) \, \psi_y - \frac{g_T - g_S}{2}\psi_x^* \psi_y^2 +  F \label{GPE1}
\end{equation}
\begin{equation}
i\partial_t \psi_y =
\left[ \omega_{LP} (\hat{\mathbf{k}}) + \frac{\alpha}{2} \cos(2\Theta) + \frac{g_T+g_S}{2} |\psi_y|^2 +  g_T |\psi_x|^2 + g_R n_R  -i \frac{\gamma_c  + \gamma_{in}}{2} \right] \psi_y - \frac{\alpha}{2} \sin(2\Theta) \, \psi_x - \frac{g_T - g_S}{2}\psi_y^* \psi_x^2
\label{GPE2}
\end{equation}
\begin{equation}
\partial_t n_R = - \gamma_R n_R + \gamma_{in} (|\psi_x|^2 + |\psi_y|^2)
\label{res}
\end{equation}
\end{widetext}
where  $\omega^{LP} (\hat{\mathbf{k}}) = \omega_0^{LP} -\frac{\hbar}{2 m} \mathbf{\nabla}^2 $ and $\hat{\mathbf{k}} = -i \nabla$, with $m$ the effective polariton mass,
$\alpha \sim 0.1 \pm 0.05$ meV is a birifringence splitting and  $\Theta \simeq 19^{\circ}$ is the angle between the natural cavity axis and the pump-probe $x,y$ polarization basis.
$g_T$ and $g_S$ are  the triplet and singlet coupling constants respectively, and we take  $g_S= -0.1 \ g_T$ and $g_T>0$.  The pump field is $F({\mathbf r}, t) = F_0 e^{i {\mathbf k}_p\cdot {\mathbf r}-i\omega_p t}$. The other parameters are the   reservoir filling rate  $\gamma_{in}$ from the condensate,  the reservoir decay rate $\gamma_R$ and the condensate radiative loss rate $\gamma_c$.

\subsubsection*{Steady-state}  If the detuning $\Delta=  \omega_p - \omega_0^{LP}$ is  sufficiently large, the steady state solution $\{\psi^s_x,\psi^s_y,n^s_R\}$ of Eqs.(\ref{GPE1}-\ref{res}) displays  a bistable behaviour for $\psi^s_x$ as a function of the pump intensity $|F_0|^2$  , similar to the well-known solution in the  single-polarization case \cite{carusotto_2013}, while for $\psi^s_y$ the upper branch of the hysteresis is decreasing, due to the competing interactions with $\psi^s_x$.

\subsubsection*{Excitation spectrum}   
\begin{figure}[bt]
\includegraphics[width=0.6\columnwidth]{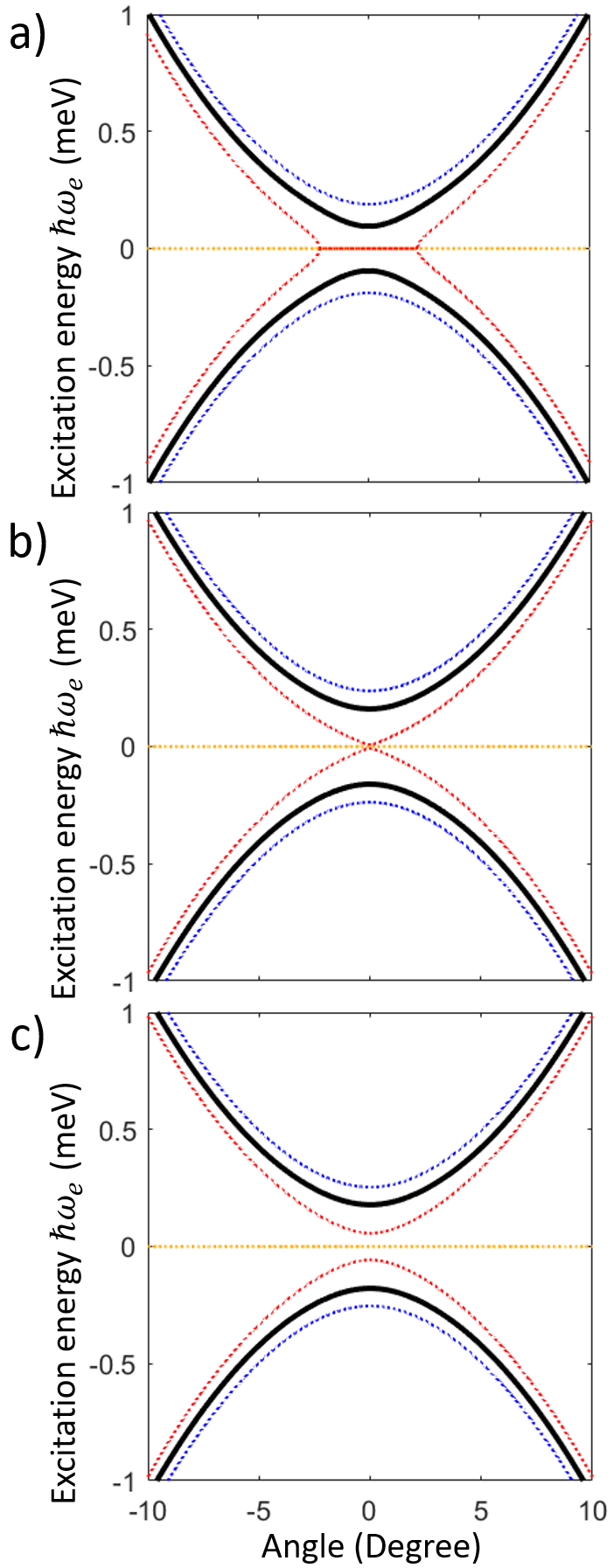}
\caption{\corr{\textbf{Dispersion relations of the excitations in the vectorial model} - The five theoretical branches obtained from Eq.(\ref{BdG}) are shown as dotted lines. The mostly reservoir excitation branch is shown in orange. The mostly polaritonic excitation branches are shown in red (dominant $x$ character) and blue (dominant $y$ character). The normal (ghost) branches have a positive (negative) energy. The black line shows the theoretical dispersion as extracted from the calculated $I(\textbf{k},\omega)$. The parameters are those of WPA: $\Delta = 0.79\,$meV, $\Theta=\pi/9$, $\alpha=0.1\,$meV, $g_S=-0.1g_T$, $\hbar\gamma_R=\hbar\gamma_i=0.0016\,$meV. Different pump strength are shown: (a) $|F|^2\simeq|F_{ju}|^2$ (a), (b) $|F|^2=1.03|F_{ju}|^2$, and (c) $|F|^2=1.05|F_{ju}|^2$, where $|F_{ju}|^2$ is the jump-up pump strength.}}
\label{figS10}
\end{figure}

By linearizing  Eqs.(\ref{GPE1}-\ref{res}) for small deviations from the steady state, ie  $\delta \psi_\sigma=\psi_\sigma-\psi^s_\sigma$, $\delta n_R=n_R-n^s_R$, we obtain the  generalized Bogoliubov-De Gennes equations
\begin{equation}
  i \partial_t \delta \vec{\psi} (\mathbf{r}, t) = \hat{\mathscr{L}}(\hat{\mathbf{k}}) \delta \vec{\psi} (\mathbf{r}, t)
  \label{BdG}
\end{equation}
where  $\delta \vec{\psi} = (\delta \psi_x, \delta \psi_x^*, \delta \psi_y, \delta \psi_y^*, \delta n_R )$ is the fluctuation vector. The Bogoliubov spectrum then consists in five eigenbranches $\omega(k)$ related by particle-hole symmetry. \corr{Fig.\ref{figS10} show examples of the five eigenbranches visible in the direction cross-polarized with the pump, calculated with the parameters of WPA, and different pump strength $|F|^2$. The flat branch has a mostly reservoir character, while the four others have a mostly polaritonic character. Owing to the brifreingence characteristic ($\alpha=0.1\,$meV and $\Theta=19^\circ$) The red modes originates mostly from the $x$ component (co-polarized with the laser drive), while the blue modes originate mostly from the $y$ component (cross-polarized with the laser drive).}

In order to account for the experimental observations, we perform  a linear-response analysis for the field amplitude $\delta \vec{\psi}= (\delta \psi_x(\mathbf{k}), \delta \psi_x^*(-\mathbf{k}), \delta \psi_y(\mathbf{k}), \delta \psi_y^*(-\mathbf{k}), \delta n_R(\mathbf{k}))$
\begin{equation}
\delta \vec{\psi} = [\omega - \hat{\mathscr{L}}(\mathbf{k})]^{-1} \delta \vec{F}
\end{equation}
as the response to a stochastic drive   $\delta \vec{F} = (\delta F_x(\mathbf{k}, \omega), -\delta F_x^*(-\mathbf{k}, -\omega), \delta F_y(\mathbf{k}, \omega), -\delta F_y^*(-\mathbf{k}, -\omega), 0 )$. Assuming that the main excitation mechanism is the coupling to acoustic phonons \cite{savenko_2013}, one has  $\delta F_{\sigma}(\mathbf{k}, \omega) = \psi^s_{\sigma}  {\cal T}({\mathbf{k}, \omega})$, with ${\cal T}({\mathbf{k}, \omega})$ a  stochastic phonon field. After averaging over the random realizations of the phonon field and defining   $\chi_{ij}( \mathbf{k}, \omega) = \left[ \frac{1}{\omega - \hat{\mathscr{L}}(\mathbf{k})} \right]_{ij}$, the  field intensity is
\begin{align}
  &\langle\langle |\delta \psi_y (\mathbf{k}, \omega)|^2 \rangle\rangle_{ph} =\nonumber \\
  &|\chi_{31} \psi^s_x - \chi_{32} {\psi^s_x}^* + \chi_{33} \psi^s_y  - \chi_{34} {\psi^s_y}^* |^2  \ S_{ph}(\mathbf{k}, \omega)
  \label{eq:chis}
\end{align}
where $S_{ph}$ is the phonon density of state, taken in this work as constant.  Finally, in order to model the effect of the finite pump spot and of the iris, we multiply the polariton field by a spatial filter function $f$ corresponding to a circular hole of diameter $35 \mu m$, $\psi_y^{out}(\mathbf{r}, t) = f(\mathbf{r}) \psi_y (\mathbf{r}, t)$; in the end, the measured Fourier space intensity is
$ I(\mathbf{k}, \omega) = \int d\mathbf{k}' |{f}(\mathbf{k} - \mathbf{k}')|^2  \langle\langle |\delta \psi_y (\mathbf{k}', \omega)|^2 \rangle\rangle_{ph}$. \corr{$I(\mathbf{k}, \omega)$ is our actual experimental observable. We thus apply to it the same numerical analysis as for the experimental one: we fit it with a single Lorentzian lineshape, in order to obtain a single theoretical dispersion relation (two if we accounts for positive and negative energy branches), to be compared with the experimental one. The result of this procedure is shown as a black line in Fig.\ref{figS10}.}

\begin{figure}[bt]
\includegraphics[width=0.6\columnwidth]{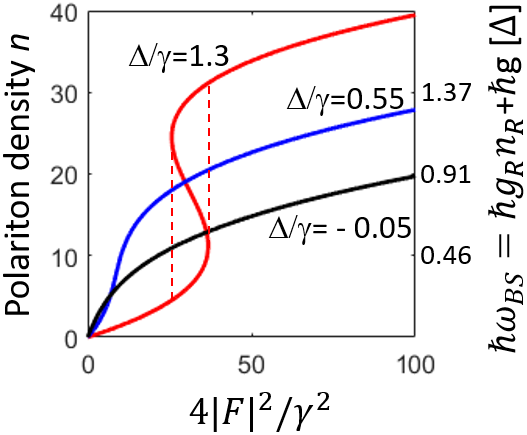}
\caption{\textbf{Hysteretic response of $n(|F|^2)$ in the scalar limit,} in three different conditions: $\Delta/\gamma=-0.05$ (black line), $\Delta/\gamma=0.55$ (blue line), and $\Delta/\gamma=1.3$ (red line). Only the latter exhibits a hysteretic behaviour. The horizontal axis is the unitless polariton generation rate normalized to the loss rate $4|F|^2/\gamma_c^2$.}
\label{figS9}
\end{figure}

\corr{\subsubsection*{Scalar limit of the model}

In the main text, we take the scalar limit of the vectorial model as it is much simpler to discuss. It would be valid for instance under circular excitations, and co-polarized detection. Its expression reads
\begin{eqnarray}
  i\hbar \partial_t \psi&=&\left[\hbar \omega_0 -\frac{\hbar^2}{2m} \nabla^2 + \hbar g|\psi|^2 + \hbar g_R n_R\right.  \nonumber
  \\ &-&\left.i \frac{\hbar (\gamma+\gamma_{in})}{2}\right]\psi +F(t)  \label{eq:cond} \\
  \partial_t n_R &=&-\gamma_R\, n_R +\gamma_{in}|\psi|^2,
  \label{eq:reserv}
\end{eqnarray}
where $m$ is the polariton effective mass, $F(t)$ is the resonant, spatially homogeneous laser drive, $\gamma$ is the radiative loss rate, and the much slower capture rate of polaritons by the reservoir is $\gamma_{in}$. The polariton-polariton interaction energy is proportional to the scalar density $n=|\psi|^2$ with a coupling constant $g$, while the interactions energy between polaritons and the reservoir is fixed by $\hbar g_R n_R$.

The corresponding steady-state solution for the polariton density reads
\begin{equation}
\left[(-\hbar\Delta+\hbar g_{\rm eff}n)^2+\frac{\hbar^2\bar{\gamma}^2}{4}\right]n=|F|^2,
\end{equation}
where $g_{\rm eff} = g + g_R\gamma_{in}/\gamma_R$ and $\bar{\gamma}= (\gamma + \gamma_{in})$, and the total blueshift $\hbar\omega_{BS}=g_{\rm eff}n$. This function exhibits a hysteretic behaviour as soon as $\hbar\Delta\leq \hbar\gamma\sqrt{3}/2$. Examples of steady-state polariton density $n$ versus pump strength $|F|^2$ are shown in Fig.\ref{figS9}.}

\subsection{Derivation of the model}

We provide here below the derivation of the Hamiltonian used to model the experiment.

\subsubsection*{Kinetic part} The cavity used in the experiment has some intrinsic birifringence. In the linear polarization basis $| s \rangle , | f \rangle$, associated to the axis of the cavity, the Hamiltonian  for the lower polariton (LP) branch reads
\begin{eqnarray}
  H_{bir} &=& \int d^2r \,\hat{\psi}_s^{\dagger}  \left(\omega^{LP}(\hat{\mathbf{k}}) - \frac{\alpha}{2} \right)    \hat{\psi}_s \nonumber \\
  &+&\hat{\psi}_f^{\dagger}   \left( \omega^{LP} (\hat{\mathbf{k}}) + \frac{\alpha}{2}\right)  \hat{\psi}_f,
\end{eqnarray}
where  the LP band is taken in the parabolic approximation $\omega^{LP} (\hat{\mathbf k}) = \omega_0^{LP} -\frac{\hbar}{2 m} \nabla^2 $, with $m$ the effective polariton mass and the field operators  $\hat{\psi}_s({\mathbf r})$, $\hat{\psi}_f({\mathbf r})$ destroy a boson with polarization $s$, $f$ respectively at spatial position ${\mathbf r}$.

In our experiment the birefringence splitting is $\alpha \sim 0.1 \pm 0.05$ meV and the laser is pumped with a linear polarization $x$ rotated by an angle  $\Theta \simeq - 19^{\circ}$ with respect to the $s$ axis, ie  we have
\begin{equation}
\begin{pmatrix}
 \hat{\psi}_s   \\
 \hat{\psi}_f
\end{pmatrix} =
\begin{pmatrix}
\cos\Theta & \sin\Theta \\
-\sin\Theta &  \cos\Theta
\end{pmatrix}
\begin{pmatrix}
 \hat{\psi}_x  \\
 \hat{\psi}_y
\end{pmatrix} ,
\end{equation}
%
so that  the kinetic part of the Hamiltonian in the  $| x \rangle , | y \rangle$ linear polarization basis reads
\begin{widetext}
\begin{equation}
H_{kin} = \int d^2r \,
\begin{pmatrix}
\hat{\psi}_x^{\dagger} &  \hat{\psi}_y^{\dagger}
\end{pmatrix}
\left[
 \omega^{LP} (\hat{\mathbf{k}})     \mathbb{I}+ \frac{\alpha}{2}
\begin{pmatrix}
-\cos2\Theta & -\sin2\Theta \\
-\sin2\Theta &  \cos2\Theta
\end{pmatrix}
\right]
\begin{pmatrix}
\hat{\psi}_x  \\
\hat{\psi}_y
\end{pmatrix} ,
\end{equation}
\end{widetext}

\subsubsection*{Two-body interactions}
Using spin conservation, the polariton--polariton interaction is naturally written  in the circular polarization basis $|\sigma_{\pm}\rangle=(| x \rangle \pm i | y \rangle)/\sqrt{2}$ as
\begin{eqnarray}
H_{int} &=& \frac{1}{2} \int d^2 r \left[ \ g_T ( \hat{\psi}_+^{\dagger} \hat{\psi}_+^{\dagger} \hat{\psi}_+ \hat{\psi}_+ + \hat{\psi}_-^{\dagger} \hat{\psi}_-^{\dagger} \hat{\psi}_- \hat{\psi}_-) \right. \nonumber \\  &+& \left. 2 g_S \hat{\psi}_+^{\dagger} \hat{\psi}_-^{\dagger} \hat{\psi}_+ \hat{\psi}_- \right]
\end{eqnarray}
where $\hat{\psi}_{\pm}({\mathbf r})$ denote the field operators for $\sigma_{\pm}$  circular polarizations respectively. In the simulations we take the values $g_S/g_T \sim -0.1$,  $g_T>0$, in agreement with known properties of
III-V semiconductor microcavities \cite{ciuti_1998, takemura_2014}.

The full vectorial  Hamiltonian for the polariton field then reads
\begin{equation}
H_0 = H_{kin} + H_{int} +
\int d^2 r \, \left[F({\mathbf r}, t) \hat{\psi}_x^{\dagger}({\mathbf r}) + F^*({\mathbf r}, t) \hat{\psi}_x({\mathbf r})\right]
\end{equation}
where $F({\mathbf r}, t)$ is the laser pump. Notice that in the experiment only the $x$-polarization component of the polariton field is pumped.

\subsubsection*{Reservoir population, incoherent losses and cavity losses}
We consider an excitonic reservoir coupled to the polariton density as in \cite{krizha18}, density coupled to the condensate. We also model the polariton radiative losses from the cavity by introducing a loss rate constant $\gamma_c$ for both polariton polarizations.

Using the above results in the mean-field approximation for the polariton fields,  one readily derives the Gross--Pitaevskii equations  given in Eq.(3) in the main text, as well as the equation for the reservoir (Eq.(4) of the main text).

\subsubsection*{Polariton-phonon interactions}
Finally, we describe the coupling of polaritons to acoustic phonons by a deformation-potential interaction via the Hamiltonian \cite{piermarocchi_1996}
\begin{equation}
H_{pol-phon} = \sum_{{\mathbf q},q_z} G_{{\mathbf q},q_z}(b_{{\mathbf q},q_z}-b^\dagger_{-{\mathbf q},-q_z})\rho_{{\mathbf q}}
\end{equation}
where $G_{{\mathbf q},q_z}$ is the acoustic phonon-polariton coupling strength taking into account the anisotropy due to the presence of the  quantum well confinement in the $z$ direction \cite{piermarocchi_1996, savenko_2013},  $b_{{\mathbf q},q_z}$ is the phononic field operator and   $\rho_{{\mathbf q}}$ is the density fluctuation of the polariton condensate with cavity in-plane momentum ${\mathbf q}$.

The above Hamiltonian yields a stochastic phonon field  in the Gross-Pitaevskii equation acting on both polarization components $\sigma=x,y$ of the polariton condensate according to $(\sum_{\mathbf q}[{\cal T}({\mathbf q},t)+{\cal T}^*({\mathbf{-q}},t)] e^{i \mathbf{q}\cdot\mathbf{r}}) \psi_\sigma({\mathbf r})$, where
\begin{equation}
  \langle\langle {\cal T}^*({\mathbf q},t){\cal T}({\mathbf q'},t')\rangle\rangle_{ph}=\sum_{q_z} |G_{{\mathbf q},q_z}|^2 \delta_{{\mathbf q},{\mathbf q}'} n(\omega_{\mathbf q, q_z})\delta(t-t'),
  \label{Sph}
\end{equation}
with $ \langle\langle \cdot \cdot \cdot \rangle\rangle_{ph}$ being the average over the noise realizations,  $ n(\omega_{\mathbf q, q_z})=1/(e^{\omega_{\mathbf q, q_z}/k_BT}-1)$, $T$ the temperature  and $\omega_{\mathbf q, q_z}$ the acoustic phonon dispersion.
Notice that the sum over $q_z$ ensures a non-zero matrix element at small $\mathbf q$ even if $|G_{{\mathbf q},q_z}|^2$ vanishes for $|{\mathbf q}|,q_z \rightarrow 0$.
The use of a constant power spectral density $S(\mathbf q, \omega)$ is equivalent to  the high temperature approximation of Eq. (\ref{Sph}).


As described in the methods section in the main text, within the linear response theory we treat the field ${\cal T}({\mathbf q},\omega)$ to order one in perturbation theory (so that $\psi_{\sigma}$ enters at order 0 in $\delta F$, with consequent decoupling of the Fourier components), and calculate the linear response of the polariton condensate to it.

\subsection{Bogoliubov excitations excluding birefringence}

We derive in this section the  analytical expression for the speed of sound given  in the main text and provide an analysis of the nature of Bogolubov excitations. For this purpose we consider the limit where the birifringence parameter $\alpha$ is set to zero.

For $\alpha = 0$ the steady state features $\psi_y^s = 0$, while $\psi_x^{s}$ is given by the standard bistability condition \cite{carusotto_2013}
with renormalized nonlinear coupling strength  $g_{\rm eff} = \frac{g_T + g_S}{2} + \frac{g_R \gamma_{in}}{\gamma_R} $ and the steady-state reservoir is given by $n_R^s = \frac{\gamma_{in}}{\gamma_R} |\psi_x^s|^2$.
As a consequence of the steady-state condition, the Bogoliubov  equations for the excitations  for the condensate with $y$ polarization are decoupled from those of the condensate with $x$ polarization and the reservoir:
\begin{eqnarray}
i\partial_t \delta \psi_x &=&
\left[ -\Delta - \frac{\hbar}{2m} \nabla^2 + 2 \bar{g} |\psi_x^s|^2    + g_R n_R^s  -i \frac{\bar \gamma}{2} \right] \delta \psi_x
\nonumber \\
&+& \bar{g} {\psi_x^s}^2 \delta\psi_x^* + g_R \psi_x^s \delta n_R  \label{BOGOX}
\end{eqnarray}
\begin{equation}
\partial_t \delta n_R = - \gamma_R \delta n_R + \gamma_{in} (\psi_x^s \delta \psi_x^* + {\psi_x^s}^* \delta \psi_x) \label{deltanR},
\end{equation}
and
\begin{eqnarray}
i\partial_t \delta \psi_y &=&
\left[ -\Delta - \frac{\hbar}{2m} \nabla^2 + g_T |\psi_x^s|^2 + g_R n_R^s  -i \frac{\bar \gamma}{2} \right] \delta\psi_y
\nonumber \\
&-& g_d {\psi_x^s}^2  \delta \psi_y^* ,
\label{BOGOY}
\end{eqnarray}
where $\Delta=\omega_p -\omega_{LP}^0$, $\bar{g} = (g_T + g_S)/2$, $g_d= (g_T - g_S)/2$  $\bar \gamma= (\gamma_c + \gamma_{in})$.
The corresponding $5 \times 5$ Bogoliubov matrix ${\mathscr{L}}(\hat{\mathbf{k}})$ separates into two block matrices, namely a 3$\times$3 part for the $x$-polarized condensate and reservoir, and a $2\times 2$ part for  the $y$-polarized condensate.

We start discussing the  3$\times$3 part:
\begin{equation}
\left. \mathscr{L} (\mathbf{k})\right|_{x,R} =
  \begin{pmatrix}
\eta_x(\mathbf{k})- i \frac{\bar \gamma}{2}  &  \bar{g} {\psi_x^s}^2 & g_R \psi_x^s \\
- \bar{g} {{\psi_x^s}^*}^2 & - \eta_x(- \mathbf{k}) - i \frac{\bar \gamma}{2} & -g_R \psi_x^* \\
i \gamma_{in} \psi_x^s& i \gamma_{in} {\psi_x^s}^* & - i \gamma_R
\end{pmatrix},
\label{L3x3}
\end{equation}
where $\eta_x(\mathbf{k}) = -\Delta +  k^2/2m + 2 \bar{g} |\psi_x^s|^2  + g_R n_R$. The structure of the above matrix is characterized  by particle-hole symmetry, ie
\begin{equation}
  \mathscr{P} \mathscr{L} = - \mathscr{L} \mathscr{P}
\end{equation}
where
\begin{equation}
\mathscr{P} = \mathscr{K}
\begin{pmatrix}
\ 0 \  & 1 \  &  0 \ \\
\ 1 \  & 0  \ & 0  \ \\
\ 0 \  & 0 \ & 1 \
\end{pmatrix}
\end{equation}
and  $\mathscr{K}$ stands for complex conjugation.
This symmetry implies $\mathscr{L} \mathscr{P} | \omega \rangle = - \omega^* \mathscr{P} | \omega \rangle$, so that $\mathscr{P}$ links pairs of eigenvectors. Since the size of the matrix is three and the system is parity invariant, this analysis allows us to immediately conclude that one eigenvlue is purely imaginary, and we attribute it to  the reservoir branch. The remaining two eigenvalues,  corresponding to the particle and hole branches, take the form $\omega_x^{\pm}(\mathbf{k}) = \pm \epsilon(k) - i \frac{\gamma(k)}{2}$.

In the case $\Delta = \hbar \omega_{BS}$, where  $\hbar \omega_{BS} = \bar{g} |\psi_x|^2   + g_R n_R $,  the excitation spectrum has a linear, gapless dispersion. By solving the eigenvalue problem at long wavelength  we obtain  $\omega_x^{\pm}(\mathbf{k}) = \pm c_x k - i \bar \gamma/2$ with
\begin{equation}
c_x^2 =  \frac{\hbar \omega_{BS}}{m} - \frac{\bar \gamma}{(\bar \gamma - 2 \gamma_R)}\frac{ g_R n_R}{m},
\label{cx}
\end{equation}
holding for $c_x^2 >0$. In the limit $\gamma_R \ll \bar \gamma = (\gamma_c +\gamma_{in})$, corresponding to the experimental conditions where the reservoir reacts slowly to fluctuations in the condensate,  we obtain  $ c_x^2 = \hbar \omega_{BS}/m - g_R n_R/m = \bar{g} |\psi_x^s|^2/m$, indicating that only the energy due to the condensate contributes to the speed of sound, while the energy stored in the dark excitonic reservoir acts just as a global energy shift.
In the adiabatic limit $ \gamma_R  \gg\bar \gamma  $, when the reservoir responds instantaneously to the condensate one finds that both reservoir and condensate contribute to the speed of sound, ie   $c_x^2 = \hbar \omega_{BS}/m$.

In the $y$-polarization sector the Bogolubov matrix reads
\begin{equation}
\left. \mathscr{L} (\mathbf{k})\right|_{y} =
  \begin{pmatrix}
\eta_y(\mathbf{k})- i \frac{\bar \gamma}{2}  &  g_d {\psi_x^s}^2 \\
- g_d {{\psi_x^s}^*}^2 & - \eta_y(- \mathbf{k}) - i \frac{\bar \gamma}{2}  \\
\end{pmatrix},
\end{equation}
where $\eta_y(\mathbf{k}) = -\Delta +  k^2/2m +  g_T |\psi_x^s|^2  + g_R n_R$.
The condition for having  gapless Bogoliubov excitations in the above equation  is $-\Delta+ g_T|\psi_x^s|^2 +  g_R n_R^s =g_d |\psi_x^s|^2$.  Interestingly, this  coincides with the condition  $\Delta=\hbar \omega_{BS}$ required for having a gapless  $x$ branch. At low momenta the dispersion relation of the $y$ branch reads  $\omega_y^{\pm}(\mathbf{k}) = \pm c_y k - i \bar \gamma/2$, with
\begin{equation}
c_y^2=\frac{g_d(\mu_y-g_R n_R)}{m}= \frac{g_T-g_S}{2}\,\frac{|\psi_x^s|^2}{m}
\label{cy}
\end{equation}
where $\mu_y = g_T |\psi_x^s|^2 + g_R n_R $.
Notice that  our choice of parameters $g_T>0$, $g_S/g_T=-0.1$  implies that the $y$ branch of the phonon dispersion lies at higher energy than the $x$ branch in all regimes, which implies that $c_y>c_x$.

\begin{figure}[htb]
\centering
\includegraphics[width = 0.45\textwidth]{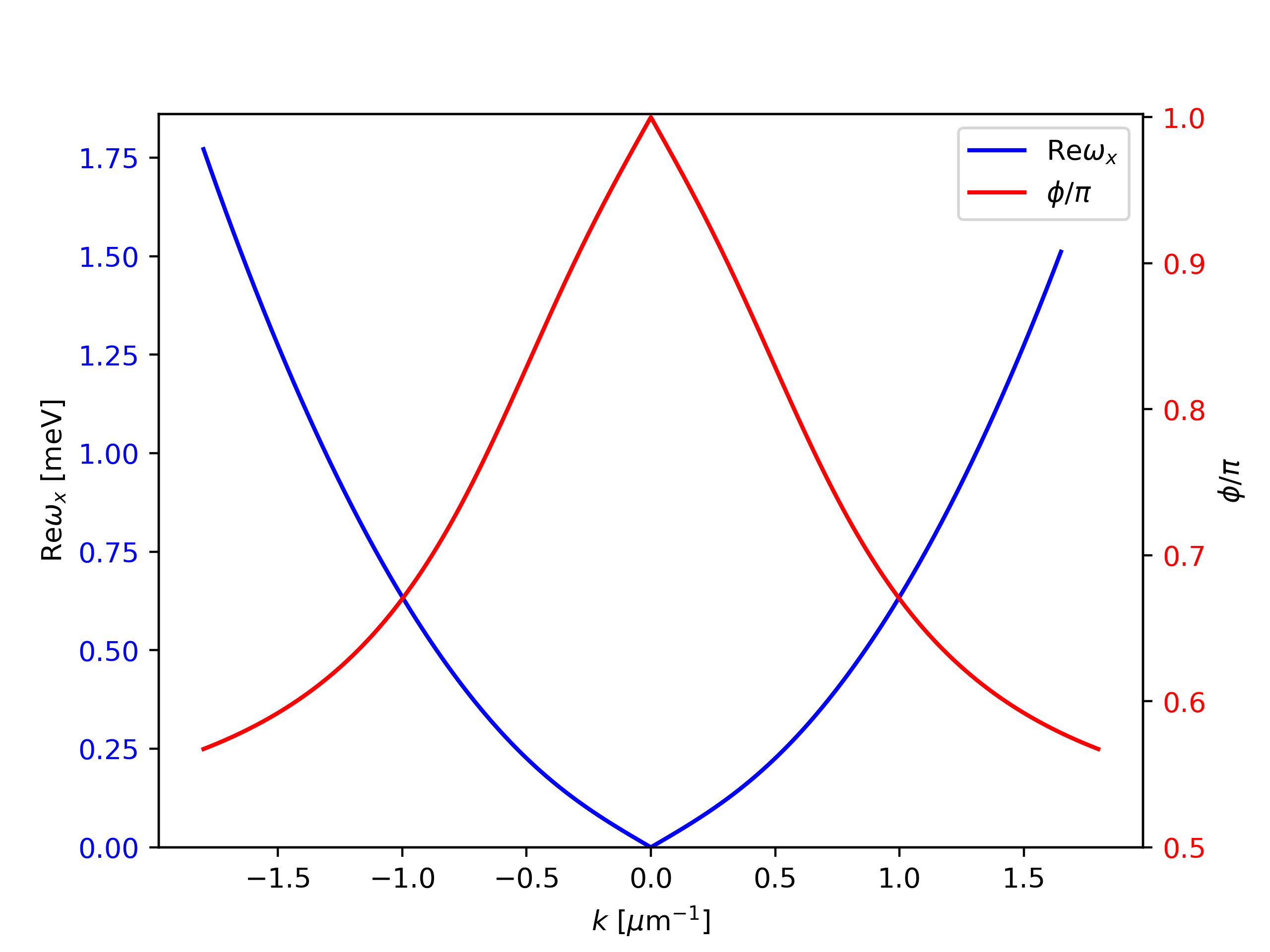}
\caption{\textbf{relative phase between the condensate and the reservoir excitations in the eigenbranch $\omega_x(k)$} - Blue, left axis: real part of $\omega_x(\mathbf{k})$. The phase difference $\phi = \arg \delta n_R / \delta |\psi_x|^2$ along the $x$ positive eigenbranch is shown in red, (right axis). For the calculation  we have chosen $\Delta =  \hbar \omega_{BS} = 0.8$ meV,  $\alpha = 0$ (so that the $y$ polaritons are decoupled), $\hbar^2/m = 1.0 $ meV$ \mu$m$^2$,  $g_R/\bar{g} = 4.0$,
   $\gamma_R = 0.0016$ meV, $\bar{\gamma} = 0.4$ meV, $E_c/E_{tot} = 0.2$. This last condition fixes $\gamma_{in} = 0.0016$ meV
}
\label{fig:nature-fasi}
\end{figure}

Next, we analyze  how the reservoir influences the nature of the eigenmodes of the Bogolubov excitations. Using  Eq.(\ref{deltanR}) above we  get the relation between the  variation of the condensate density and the one of the reservoir density,
\begin{equation}
\frac{\delta n_R }{\delta ( |\psi_x|^2 + |\psi_y|^2)  } = \frac{\gamma_{in}}{\gamma_R - i \omega_a(\mathbf{k})}, \label{eq:nature-fasi}
\end{equation}
that holds for any eigenmode. Fig. \ref{fig:nature-fasi} shows the argument of this quantity for the case of the $a=x$ branch, at $\alpha = 0$ and at the point with gapless excitation spectrum. We notice that at small wavevector $|$Re$\omega_x(k)| \ll |$Im$\omega(k)| \simeq \bar{\gamma}$ the density fluctuations of the condensate and of the reservoir  are in phase opposition. This is in agreement with the fact that
the excitation branch $\omega_x(\mathbf{k})$ is a Goldstone mode, so at small $k$ the system tries to keep constant density by   making $\delta n_R$ and $\delta |\psi_x|^2$ oscillate with a relative phase $\pi$ to compensate each other. At large momentum $|$Re$\omega_x(k)| \gg |$Im$\omega(k)|$ instead, $\delta n_R$ follows $\delta |\psi_x|^2$ in  quadrature of phase, with the condensate density fluctuations driving  the reservoir density ones. Given the complex nature of $\omega_a(\bf{k})$, the transition between the two regimes occurs when the real part of the Bogoliubov energy $\textrm{Re}[\omega_a(\bf{k})]$ is of the order of the loss rate $\bar{\gamma}$.

Finally, we extend the above analysis to the case of  finite in-plane momentum of excitation ie for  $\mathbf{k_p}\neq 0$. The steady-state solution takes the form  $\psi_x^s(\mathbf{r}) = |\psi_x^s| \ e^{i \mathbf{k_p} \cdot \mathbf{r}}$.
In the case $\alpha = 0$ the Bogoliubov matrix Eq.(\ref{L3x3}) for the $x$-polarization condensate  coupled to the reservoir is modified to \cite{carusotto_2004}
\begin{align}
& \left. \mathscr{L}_{\mathbf{k_p}} (\mathbf{k})\right|_{x,R} = \nonumber \\
&  \begin{pmatrix}
\eta_x(\delta \mathbf{k} + \mathbf{k_p})- i \frac{\bar \gamma}{2}  &  g {\psi_x^s}^2 & g_R \psi_x^s \\
- g {{\psi_x^s}^*}^2 & - \eta_x( \delta \mathbf{k} - \mathbf{k_p}) - i \frac{\bar \gamma}{2} & -g_R {\psi_x^s}^* \\
i \gamma_{in} \psi_x^s& i \gamma_{in} {\psi_x^s}^* & - i \gamma_R
\end{pmatrix}.
\end{align}

where $\delta \mathbf{k} = \mathbf{k} - \mathbf{k_p}$ is the momentum measured in the fluid reference frame.

Interestingly, in this case the inversion symmetry is broken, and the reservoir excitation branch acquires a real part (while particle--hole symmetry only requires $\mathscr{P} | \delta \mathbf{k}, \omega \rangle = | -\delta \mathbf{k}, - \omega^* \rangle $).
In particular, at the equilibrium point $\Delta - \frac{\hbar^2k_p^2}{2m} - \hbar \omega_{BS} = 0 $, the real part of the reservoir eigenvalue is of order three in $|\delta \mathbf{k}|$ for small $|\delta \mathbf{k}|$, while $\omega_x(\mathbf{k}, \mathbf{k}_p)$ in the same regime is modified to  first order in $|\delta \mathbf{k}|$ only by the  usual Doppler shift \cite{carusotto_2004}: at given $\hbar \omega_{BS}$ we have $\omega_x(\mathbf{k}, \mathbf{k}_p) =  \frac{\mathbf{k}_p \cdot \delta \mathbf{k}}{m} + \omega_x(|\delta \mathbf{k}|, 0) $.